\documentclass[twocolumn]{pasj01}
\begin{document}

\title{X-ray and Optical Observations of the Black Hole Candidate MAXI J1828$-$249}

\author{Sonoe \textsc{Oda},\altaffilmark{1,2,}$^{*}$
Megumi \textsc{Shidatsu},\altaffilmark{3,4}
Satoshi \textsc{Nakahira},\altaffilmark{3}
Toru \textsc{Tamagawa},\altaffilmark{1,2}
Yuki \textsc{Moritani},\altaffilmark{5,6}
Ryosuke \textsc{Itoh},\altaffilmark{7,8}
Yoshihiro \textsc{Ueda},\altaffilmark{9}
Hitoshi \textsc{Negoro},\altaffilmark{10}
Kazuo \textsc{Makishima},\altaffilmark{3,5} 
Nobuyuki \textsc{Kawai},\altaffilmark{3,7}
and Tatehiro \textsc{Mihara}\altaffilmark{3}}
\altaffiltext{1}{RIKEN Nishina Center, 2-1 Hirosawa, Wako, Saitama 351-0198, Japan}
\altaffiltext{2}{Department of Physics, Tokyo University of Science, 3-1 Kagurazaka, Shinjuku-ku, Tokyo 162-8601, Japan} 
\altaffiltext{3}{MAXI team, RIKEN, 2-1 Hirosawa, Wako, Saitama 351-0198, Japan} 
\altaffiltext{4}{Department of Physics, Ehime University, 2-5 Bunkyocho, Matsuyama, Ehime 790-8577, Japan} 
\altaffiltext{5}{Kavli Institute for the Physics and Mathematics of the Universe (WPI), The University of Tokyo, 5-1-5 Kashiwanoha, Kashiwa, Chiba 277-8583, Japan}
\altaffiltext{6}{Hiroshima Astrophysical Science Center, Hiroshima University, Higashi-Hiroshima, Hiroshima 739-8526, Japan}
\altaffiltext{7}{Department of Physics, School of Science, Tokyo Institute of Technology, 2-12-1 Ohokayama, Meguro, Tokyo 152-8551, Japan}
\altaffiltext{8}{Department of Physical Science, Hiroshima University, Kagamiyama 1-3-1, Higashi-Hiroshima 739-8526, Japan}
\altaffiltext{9}{Department of Astronomy, Kyoto University, Kitashirakawa-Oiwake-cho, Sakyo-ku, Kyoto 606-8502, Japan}
\altaffiltext{10}{Department of Physics, Nihon University, 1-8-14 Kanda-Surugadai, Chiyoda-ku, Tokyo 101-8308, Japan}
\email{sonoe.oda@riken.jp}


\maketitle

\begin{abstract}
We report results from X-ray and optical observations of the Galactic black hole candidate MAXI J1828$-$249, 
performed with Suzaku and the Kanata telescope around the X-ray flux peak in the 2013 outburst. 
The time-averaged X-ray spectrum covering 0.6--168 keV was approximately characterized 
by a strong multi-color disk blackbody component with an inner disk 
temperature of $\sim$0.6 keV, and a power-law tail with a photon index of $\sim$2.0. 
We detected an additional structure at 5--10 keV, which can be modelled 
neither with X-ray reflection on the disk, nor relativistic broadening of the disk emission. 
Instead, it was successfully reproduced with a Comptonization of disk photons by thermal
electrons with a relatively low temperature ($\lesssim$10 keV). We infer that the source 
was in the intermediate state, considering its long-term trend in the hardness intensity 
diagram, the strength of the spectral power-law tail, 
and its variability properties. The low-temperature Comptonization component 
could be produced in a boundary region 
between the truncated standard disk and the hot inner flow, 
or a Comptonizing region that somehow developed above the disk surface. 
The multi-wavelength spectral energy distribution suggests that the optical and UV fluxes 
were dominated by irradiated outer disk emission. 
\end{abstract}

\section{Introduction}
Black hole binaries (BHBs), consisting of a star and a stellar-mass black hole, 
are known to show various ``states'' with different X-ray properties 
(see e.g., \cite{Done2007} for reviews). The two most canonical 
ones are the ``high/soft state'' and ``low/hard state''. 
The former is generally seen at high mass accretion rates, where the 
X-ray spectrum is dominated by soft X-rays produced as multi-temperature 
blackbody radiation from a standard accretion disk (\cite{Shakura1973}). 
Many previous studies showed that the inner disk radius is kept fairly 
constant during the high/soft state (e.g., \cite{Makishima1986}; \cite{Ebisawa1993}; 
\cite{Steiner2010}; \cite{Shidatsu2011}); this suggests 
that the standard disk stably extends down to the innermost 
stable circular orbit (ISCO). 
The low/hard state is observed at relatively low mass accretion rates, where 
the X-ray spectrum is approximated by a power-law model with a photon 
index of $<$2, often accompanied by an exponential cutoff at 
$\sim$100 keV. It has been suggested that the standard disk in the low/hard state 
is truncated at a radius considerably larger than the ISCO, and turns into 
a radiatively inefficient hot accretion flow toward the 
vicinity of the central black hole (\cite{Esin1997}; \cite{Makishima2008}; 
\cite{Tomsick2009}; \cite{Shidatsu2011}, \yearcite{Shidatsu2013}).
The hard power-law component is regarded as a result of 
inverse Compton scattering between the soft X-rays from the 
truncated standard disk, and hot electrons in 
some optically-thin zones, including in particular the regions 
inside the truncation radius, of the accretion flow (e.g., 
\cite{Poutanen1998}; \cite{Zdziarski1998}; but see 
\cite{Reig2003}; \cite{Markoff2005} for interpretations with 
jets).
Properties of short-term variability 
also depend on the spectral states.
In the low/hard state, BHBs show strong X-ray intensity 
variations, of which the power spectra usually consist of 
``band-limited noise'' from $\sim 0.1$ Hz 
to 10--100 Hz, sometimes accompanied by 
quasi periodic oscillations at its high frequency edge
(see e.g., \cite{Axelsson2018}). These strong 
variations are suppressed in the high/soft state, wherein  
the standard disk is considered to fully develop down to the ISCO.

Even though previous observations in these two states 
have provided rich information, 
the structure of the black hole accretion flows and its
evolution are not yet fully understood. One of the 
long-standing mysteries is the origin of the power-law  ``hard tail'' 
seen above $\sim$10 keV in the high/soft state spectra. 
Usually, it has a photon index of about 
$\gtrsim$2.1 \citep{McClintock2006}, which is larger (steeper) 
than that of the X-ray continuum in the low/hard state, and does not 
show an exponential cut-off. If the hard tail is also produced 
by Comptonization of the photons from the standard disk,   
the Comptonizing electrons are suggested to be much hotter 
than those in the low/hard state, 
or have a non-thermal distribution (e.g., \cite{Gierlinski1999}). 
We do not either know whether the Comptonized continuum
in the low/hard state 
continuously evolves into the hard tail in 
the high/soft state, or they are of different origins, 
and if the former is the case,  
how one evolves into the other and what drives that evolution. 
Observations in the transitional phases between these 
two states, collectively called the intermediate state,  
may provide clues to these questions.

As a system evolves from the low/hard state to the 
intermediate state, 
the power-law component in the hard X-ray band steepens, 
and the disk blackbody component gradually 
increases its fraction to the total X-ray flux. This is often 
explained by changes in the disk truncation radius; as 
the disk develops inwards and thereby the number 
of seed photons increases, the Comptonizing hot electrons 
are more efficiently cooled (e.g., \cite{Done2007}; \cite{Sobolewska2011a}).
However, how the physical structure of the Comptonization 
component evolves through in the intermediate state, 
and how it is connected to the hard tail in the high/soft state, 
are still unclear. In addition, the stability and typical 
duration of the intermediate state
remain as another open question.

Observing the intermediate state is also important 
in understanding the physical mechanisms of outflows.
The emission from compact jets seen in the low/hard state 
is suppressed after the transition to the high/soft state,
whereas the absorption lines from disk winds 
behave in opposite ways \citep{Ponti2012}.
This indicates that these outflows are also 
state-sensitive phenomena. However, the exact 
timing of their appearance/disappearance is 
not yet clear. The intermediate state may also provide 
a key to understanding the hysteresis of the state transition; 
the transition from the low/hard state 
to the high/soft state (the hard-to-soft transition) 
usually occurs 
at a considerably higher luminosity than the transition 
in the reverse direction 
(the soft-to-hard transition). 
Thus, through their outbursts, BHBs draw counter-clockwise, 
``q''-shaped tracks in their hardness-intensity diagrams 
(\cite{Homan2005}), although different tracks are seen 
in some cases (e.g., \cite{Nakahira2014}).

\begin{figure*}
	\begin{center}
		\FigureFile(75mm, 75mm){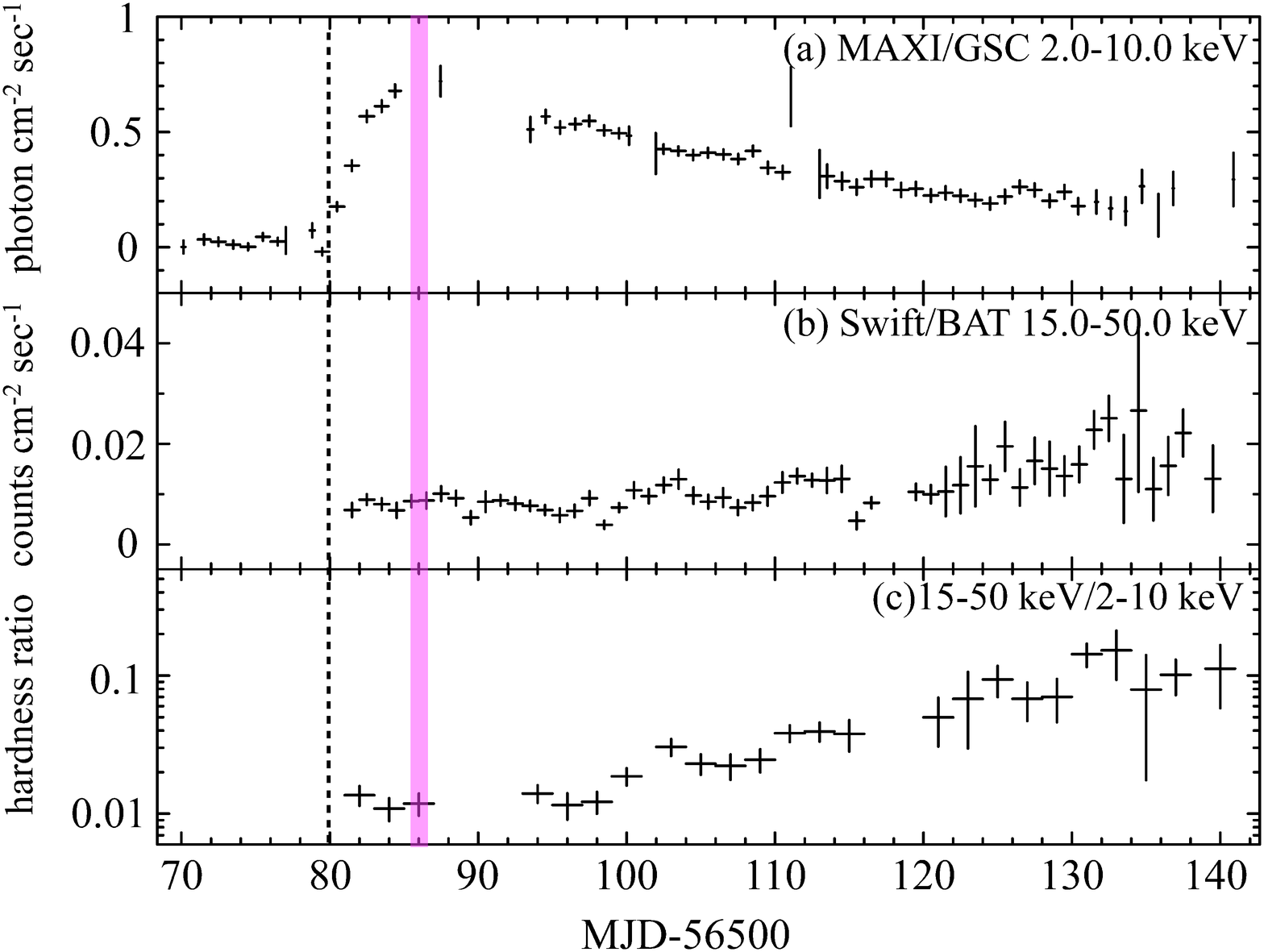} 
		\FigureFile(75mm, 75mm){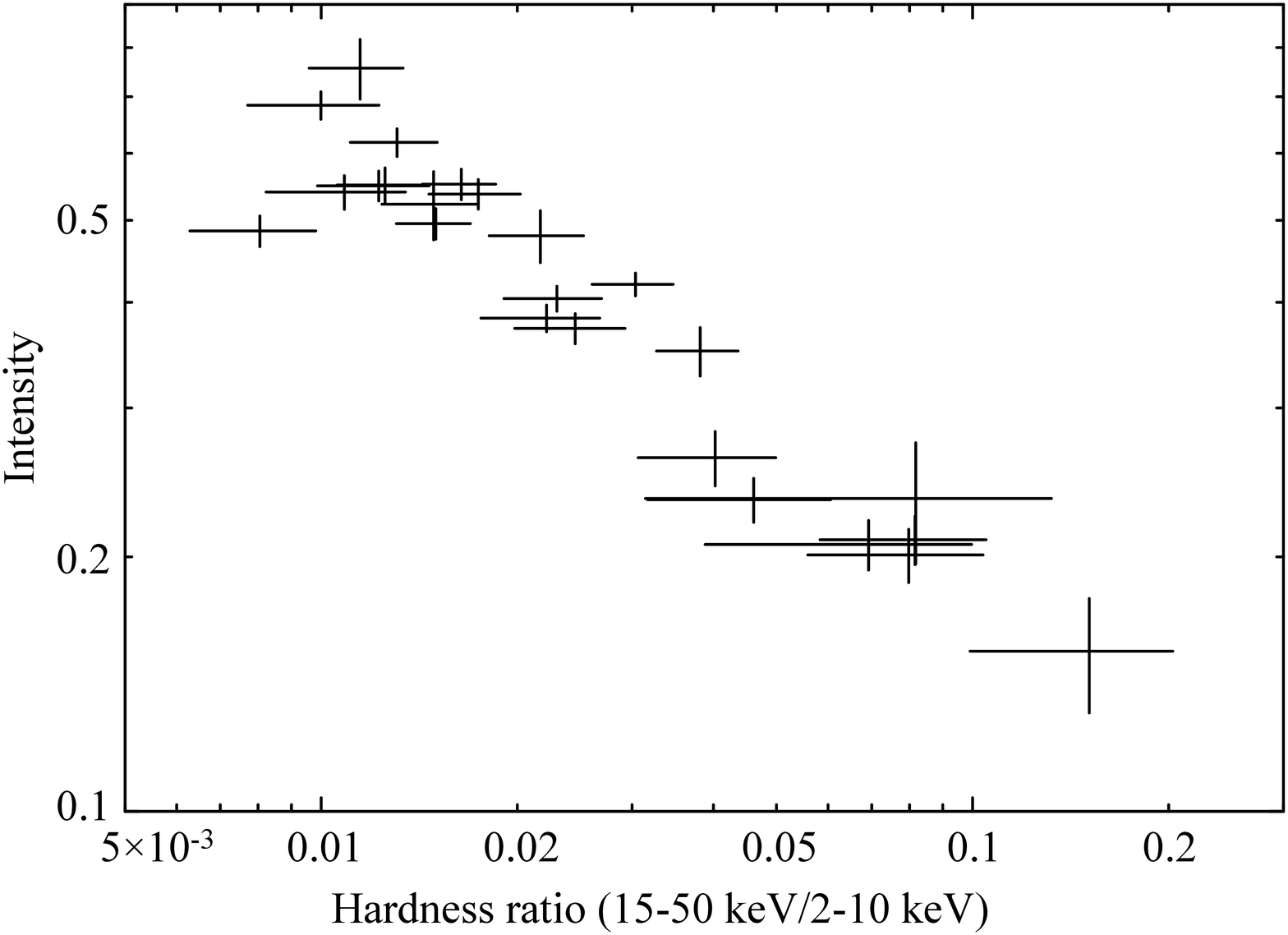} 
	\end{center}
\caption{Left: Long-term light curves of MAXI J1828$-$249 in 2--20 keV (panel a) and 15--50 keV (panel b), 
obtained with the MAXI/GSC and the Swift/BAT, respectively, 
and the hardness ratio between the 2--10 keV and 15--50 keV 
bands (panel c). Our Suzaku observation is indicated as the magenta stripe. 
The dashed line indicates the first MAXI detection and 
the start of spectral softening.
MJD 56580 corresponds to 2013 October 15. Right: A hardness-intensity diagram created 
from the MAXI/GSC and Swift/BAT light curves over MJD 56580--56640. 
}
\label{fig:fig1}
\end{figure*}

MAXI J1828$-$249 is a BHB candidate discovered with 
the Gas Slit Camera (GSC; \cite{Mihara2011}) onboard MAXI 
(Monitor of All-sky X-ray Image; \cite{Matsuoka2009}),  
on 2013 October 15 UT 21:55 (\cite{Nakahira2013}), through 
the MAXI nova alert system \citep{Negoro2016}. 
Since its discovery, the source was extensively observed with 
MAXI, as well as Swift 
(\cite{Gehrels2004}) and INTEGRAL (\cite{Winkler2003}). 
The precise position was determined with the Swift/X-ray Telescope 
(XRT; \cite{Burrows2005}) at ($\alpha^{2000}$, 
$\delta^{2000}$) 
$=$ (18$^\mathrm{h}$28$^\mathrm{m}$58.08$^\mathrm{s}$, 
$-25^\circ$ 01$'$ 45$''$.3)
(\cite{Kennea-location2013}, \cite{Negoro2016}). 
A spectral softening was observed for several days 
after the MAXI discovery \citep{Negoro2013, Kennea2013, Krivonos2013}. 
The optical and near-infrared counterpart was  
detected in that period, with apparent magnitudes of 
$\sim$17 mag in the AB magnitude system, and 
the spectral energy distribution in the optical 
and near-infrared band were successfully
explained by an outer disk emission \citep{Rau2013}.
On 2013 October 21, 6 days after the 
spectral softening was first detected, we 
triggered a target-of-opportunity (ToO) observation 
of MAXI J1828$-$249 with Suzaku (\cite{Mitsuda2007}), 
aiming at obtaining high-quality X-ray data of the target. 
We also carried out optical photometric observations 
with the Kanata telescope quasi-simultaneously with 
the Suzaku observation, to constrain 
the basic parameters of this system, such as the 
black hole mass, the distance, the type of the 
companion star, and the size of 
the accretion disk. 

In the present paper, we describe the MAXI and Suzaku observations 
of MAXI J1828$-$249 (Section~\ref{sec:Xobs}), analyze the Suzaku data 
(Section~\ref{sec:Xana}), and study broad-band spectral energy 
distribution (Section~\ref{sec:OPTobs}).
Throughout the article, errors represent 90\% 
confidence ranges, unless otherwise stated.  
We adopted the solar abundance table given by \citet{Wilms2000}.

\section{X-ray Observations}\label{sec:Xobs}

\subsection{Long-term X-ray evolution in the outburst} \label{sec:trend}

Figure~\ref{fig:fig1} shows the long-term X-ray light curves of 
the source in 2--20 keV and 15--50 keV, obtained with the 
MAXI/GSC and the Swift/BAT, respectively, and the 
hardness ratio between them. 
After the discovery on 2013 October 15, the soft X-ray flux 
rapidly increased, and reached a peak in several days. 
In the meantime, the X-ray spectrum kept softening 
as detected with MAXI, Swift, and INTEGRAL
(\cite{Kennea2013}; \cite{Negoro2013}; \cite{Krivonos2013}).
It is hence suggested that the source was evolving through  
the low/hard state toward the high/soft state, even though 
a transition to the genuine high/soft state was not clearly recognized.
X-ray spectra obtained in this period were characterized 
with a strong multi-color disk blackbody component and a 
power-law tail with a photon index of $\sim$2, seen 
mainly below and above $\sim$10 keV, respectively  
\citep{Filippova2014,Grebenev2016}. 
Then, the flux gradually decreased for $\sim$100 days 
down to undetectable levels. Meanwhile, the hardness ratio gradually 
increased, and then reached a maximum around MJD 56619, 
suggesting that the source returned to the low/hard state.
Indeed, \citet{Tomsick2014} observed a power-law shaped spectrum, 
typical of the low/hard state, in 
a Swift observation performed $\sim$4 months after 
the suggested first transition. 
Radio emission was not detected on MJD 56583, 
in the period of the X-ray spectral softening \citep{Miller2013}, 
but detected on MJD 56704, after the spectral hardening 
\citep{Corbel2014}. As shown in the right panel of 
figure~\ref{fig:fig1}, the source exhibited no significant 
hysteretic behavior in its hardness intensity diagram.

\subsection{Suzaku Observation and Data Reduction} \label{sec:suzaku_red}
The Suzaku observation of MAXI J1828$-$249 was performed 
from 2013 October 21 UT 05:08:22 to October 22 UT 05:00:11 
(OBSID$=$908002010). This was 6 days after the beginning of 
the spectral softening described in subsection 2.1. 
As shown in the left panel of figure~\ref{fig:fig1},
we caught 
the source in a brightest and softest phase in its 2013 outburst.

\begin{figure}
	\begin{center}
		\FigureFile(75mm, 75mm){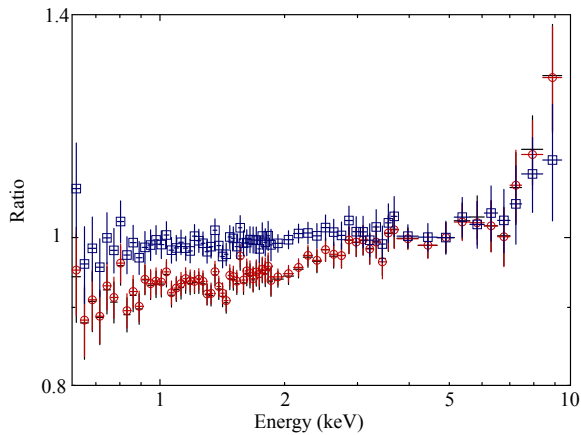} 
	\end{center}
\caption{Ratios of the background-subtracted XIS0 spectra 
extracted from a whole circular region ($r < 108''$, black crosses), the 
$r=5''.6 - 108''$ annulus (red circles; with the pileup fraction of $<$3\%), and 
the $r=44'' - 108''$ annulus (blue squares; with $<$1\% pileup fraction), 
all divided by that from the $r=70'' - 108''$ annulus.
They are all normalized to unity at 5.0 keV.
}
\label{fig:spec_ratio}
\end{figure}

Suzaku carries two X-ray instruments: the X-ray Imaging Spectrometer 
(XIS; \cite{Koyama2007}) and the Hard X-ray Detector (HXD; \cite{Takahashi2007}).
The XIS is X-ray CCD cameras covering the soft X-ray band in the 
0.2--12 keV range. In our observations, two cameras using front-side-illuminated 
chips (FI-XISs: XIS0 and XIS3) and one camera with back-side-illuminated configuration  
(BI-XIS: XIS1) were available. 
The HXD is a non-imaging, collimated hard X-ray detector 
composed of silicon PIN diodes and gadolinium silicon oxide (GSO) 
crystal scintillators, sensitive to 10--70 keV and 40--600 keV, 
respectively. 
Our observation was carried out in the XIS nominal position. 
The XIS was operated in the 1/4 window mode 
with the 0.1 s burst option, 
to avoid pile-up effects. The net exposure 
of the XIS was $\approx$2.5 ks, and those of PIN and GSO were $\approx$32 ks. 
The much shorter exposure of the XIS is due to the burst option.

The data reduction and analysis were carried out with
HEAsoft version 6.19 and Calibration Database released on 2016 April 24. 
We started the data reduction 
from the ``cleaned'' event files processed with the 
pipeline version 1.1.0, and followed the 
standard procedure described in the Suzaku 
data analysis manual\footnote{https://heasarc.gsfc.nasa.gov/docs/suzaku/analysis/abc/}. 
Source photons in the XIS data were extracted 
from a circular region with a radius of $108''$, 
centered on the target position.
Following \citet{Yamada2012}, we excluded the core of the point spread function 
with a radius of $44''$, so that the pile-up fraction was reduced to $<$1\%. 
In fact, as shown in figure~\ref{fig:spec_ratio}, the source spectrum 
accumulated over an annulus of $r = 44'' - 108''$ ($r$ being the radius) has almost the same shape 
as that from a larger annulus of $r=70'' - 108''$ (ratios in blue), although those from 
smaller inner radii (red for $r = 5''.6 - 108''$ and 
black for $r = 0'' - 108''$) 
exhibit shape distortion due to the pileup.
Compared with the spectrum from the largest annulus of 
$r = 70'' - 108''$, that from the $r = 44'' - 108''$ 
annulus still shows a slight difference above $\sim 7$ keV, 
which could be caused by the radial dependence of the 
response. This difference, however, has been confirmed 
not to affect the conclusion of this article. 
Background events were taken from a 
circle with a $90''$ radius in a blank-sky area. We have 
confirmed that the results presented in the following 
sections are unaffected by the choice of the position, size, and shape 
of the background region. We generated the response matrix files 
(RMFs) and ancillary response files (ARFs) of the individual 
XIS chips, using the HEAsoft tool {\tt xisrmfgen} and 
{\tt xissimarfgen}, respectively. 

Using the tools {\tt hxdpinxblc} and {\tt hxdpinxbpi}, 
we created a light curve and a spectrum of the PIN background, which consists 
of the cosmic X-ray background (CXB) and non X-ray background (NXB). The light curve 
and spectrum of the GSO background were made with {\tt hxdgsoxblc} and {\tt hxdgsoxbpi}, respectively. 
We ignored the CXB of GSO, because it contributes 
only 2~\% of the total background. 
These products were obtained with the modelled NXB data provided by the Suzaku 
team\footnote{http://www.astro.isas.jaxa.jp/suzaku/analysis/hxd/pinnxb, gsonxb 
(for the PIN and GSO NXB data, respectively)}. 
We adopted ae\_hxd\_pinxinome11\_20110601.rsp to generate the RMF of PIN, 
and ae\_hxd\_gsoxinom\_20100524.rsp 
and ae\_hxd\_gsoxinom\_crab\_20100526.arf
for the RMF and ARF of GSO, respectively. 

To assess the reproducibility of the NXB model of 
GSO, its prediction was compared with the data taken during 
Earth occultations of the target source in the actual observation. 
Then, in terms of the time averaged spectrum, as well as a 
128 s bin light curve in 50--100 keV and that in 100--200 keV, 
the NXB model prediction agreed with the Earth occultation data,  
within the 1$\sigma$ systematic errors 
given in \citet{Fukazawa2009}. 
Specifically, the model-to-data ratios of the count rates, 
calculated every 128-s time interval, were found at 
$0.995 \pm 0.006$ and $1.000 \pm 0.005$ 
(where the errors represent 
1$\sigma$ standard errors over 143 data points) 
in 50--100 keV and 100--200 keV, respectively. Similarly, when the time 
averaged spectrum of the NXB model is compared with that of the occultation 
data, channel-by-channel ratios between the two spectra were obtained as 
$0.997 \pm 0.006$ in 50--100 keV and $0.998 \pm 0.005$ in 
100--200 keV. 
Since the NXB model was thus confirmed to be reliable, 
and since it has higher statistics than the Earth occultation data, 
we selected the modeled NXB in subtracting the HXD-GSO background.

\begin{figure*}[h]
	\begin{center}
		\FigureFile(150mm, 150mm){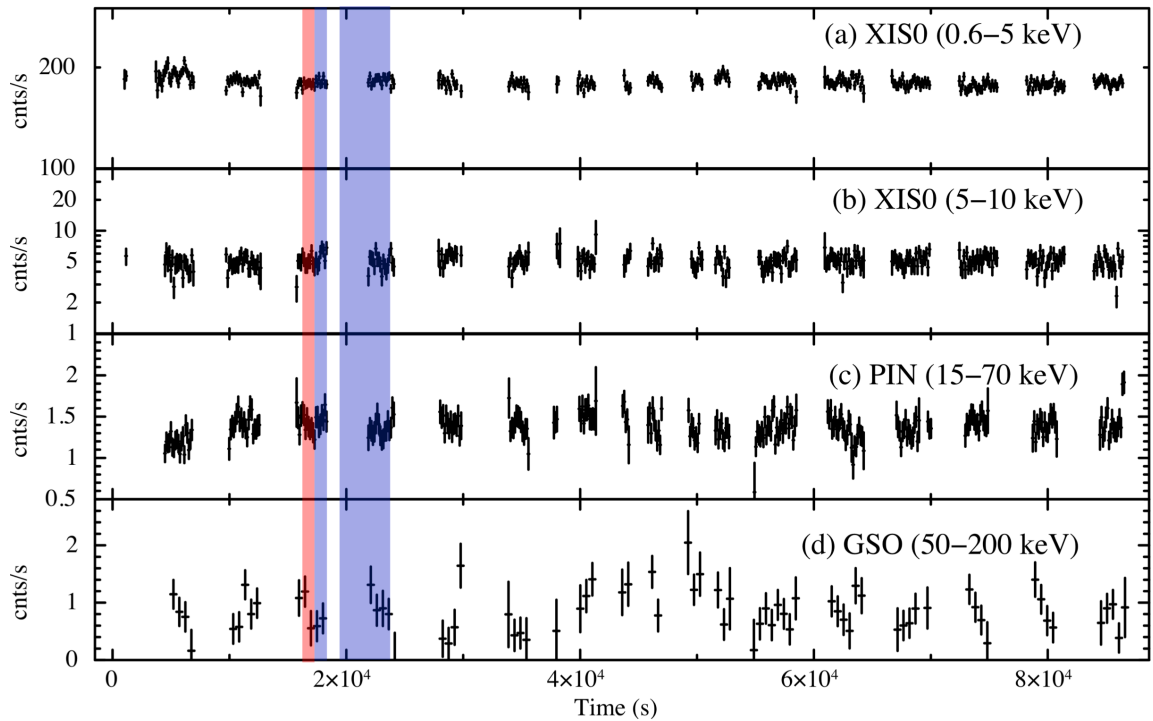}
	\end{center}
\caption{
Background-subtracted light curves of MAXI J1828$-$249 in 0.6--5 keV (panel a, with XIS0), 5--10 keV (panel b, the same), and  
15--70 keV (panel c, HXD-PIN) with 128-s bins, and that in 50--200 keV (panel d, HXD-GSO) with 512-s bins. The HXD data were corrected for dead time. 
A red shaded region indicates the period of our optical $I$-band observations 
with the Kanata telescope, and blue regions those of the $R$-band observations.
The XIS light curves employ logarithmic scales, whereas the HXD ones linear scales.
}
\label{fig:fig2}
\end{figure*}

\section{Analysis and Results of the Suzaku Data} \label{sec:Xana}
\subsection{Light Curves and Spectra}
\begin{figure}[h]
	\begin{center}
		\FigureFile(75mm, 75mm){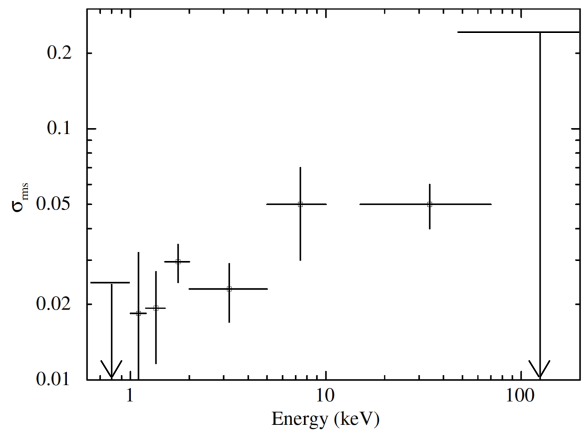}	
	\end{center}
\caption{Energy dependence of the rms variability $\sigma_\mathrm{rms}$ obtained from the 
XIS0, PIN, and GSO light curves in 128 s bins.}
\label{fig:rms}
\end{figure}

Figure~\ref{fig:fig2} shows the background-subtracted Suzaku light curves of 
MAXI J1828$-$249 in 0.6--5 keV, 5--10 keV, 
15--70 keV, and 50--200 keV. The former three light 
curves are binned in 128-s intervals, while the last 
one in 512 s bins to improve statistics. 
To quantitatively investigate 
intensity variation and its energy dependence, 
we calculated the excess variance $\sigma^2_{\rm rms}$, 
defined as \citep{Vaughan2003a}
\begin{equation}
 \sigma_{\rm rms}^2 = \frac{1}{N\bar{c}^2} \sum_{i=1}^N[(c_i - \bar{c})^2-\sigma_i^2],
\end{equation}
where $N$ is the total number of the light curve points, 
$\bar{c}$ is the mean count rate after the background subtraction, 
and $c_i$ and $\sigma_i$ are the count rate and its error at the 
$i$-th data point, respectively. 
Figure~\ref{fig:rms} presents an energy versus
$\sigma_{\rm rms}$ plot, obtained from 128-s bin light curves in 
different energy bands within 0.6--200 keV. The $\sigma_{\rm rms}$ 
value is almost constant at $\sim 0.05$ in 5--70 keV, but 
it slightly decreases below 5 keV, to $\sigma_{\rm rms} =$ 0.02--0.03.
Thus, on time scales of 128 s to several tens ks, 
the source varied by $\sim 5$\% in 5--70 keV, whereas 
2--3\% at lower energies. The GSO data 
in 50--200 keV provide only a weak upper limit as 
$\sigma_{\rm rms} < 0.24$. 
Although the GSO light curve exhibits some hint of variation on time scales 
of $\sim$10 ks, possibly correlated with systematic 
background changes, it is still within the statistical errors.
Because of insufficient photon statistics of the HXD and 
the limited time resolution of the XIS, 
we were unable to extract any useful information 
on variations on shorter timescales of $\gtrsim$0.1 Hz, 
where BHBs most prominently show their characteristic variations.

\begin{figure}[h]
	\begin{center}
		\FigureFile(75mm, 75mm){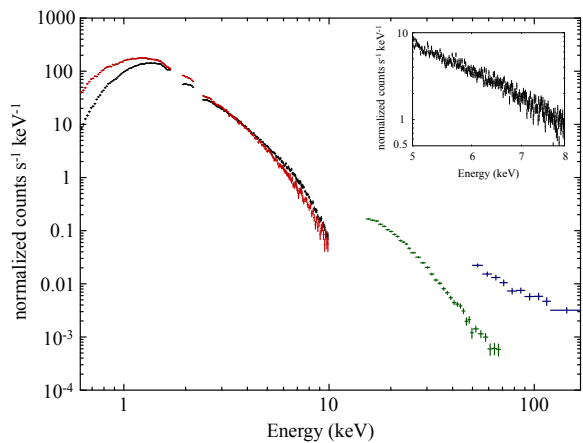}	
	\end{center}
\caption{Time-averaged spectra of MAXI J1828$-$249 obtained with XIS0 and XIS3 (black), 
XIS1 (red), PIN (green), and GSO (blue). 
The inset shows an enlarged view of the XIS0$+$XIS3 data around the iron K$\alpha$ line energies, 
obtained without discarding the image core.}
\label{fig:fig3}
\end{figure}

Figure~\ref{fig:fig3} plots the time-averaged and background-subtracted 
spectra of XIS0--3, PIN, and GSO, derived over energy ranges of 
0.6--10 keV, 15--70 keV, and 50--168 keV, respectively. 
For GSO, we employed the energy range where the source 
signals exceed 2\% of the background level, considering the systematic 
uncertainty of the GSO background (subsection 2.2; \cite{Fukazawa2009}).
The data of the two FI-XIS cameras were combined in the following 
spectral analysis to improve statistics. We excluded the XIS 
data in 1.7--1.9 keV and 2.2--2.4 keV, because of the residual  
calibration uncertainties near the instrumental Si-K edge and Au-M edge, respectively. 
In addition, to absorb calibration uncertainties in the responses, 
we included 2\% systematic errors in the individual bins of the XIS and the HXD.

\subsection{Analysis of the Hard X-ray Spectrum} \label{sec:fit_HXD}

Below, the time-averaged Suzaku spectra are analyzed in detail. 
We first focused on the HXD data, to study 
the spectral profile in the hard X-ray band 
above 10 keV. The time-averaged HXD (PIN$+$GSO) spectrum 
spanning $15$--$168$ keV was reproduced well by a power-law 
model with a photon index of $\Gamma=2.0 \pm 0.4$, 
which gave a reduced chi-squared 
of ${\chi}^2/ \rm{d.o.f}= 37/44$. The same spectrum was also 
fitted with the thermal Comptonization model {\tt nthcomp} 
(\cite{Zdziarski1996}; \cite{Zycki1999}), which computes a 
thermal Comptonization spectrum using the photon index 
$\Gamma$, the electron temperature $kT_\mathrm{e}$, 
the seed photon temperature $kT_\mathrm{bb}$, and the 
normalization as free parameters. 
We adopted the disk blackbody as the seed 
photon source, and set $kT_\mathrm{bb} = 0.64$ keV, 
which was determined by incorporating 
the XIS data (see Section~\ref{sec:double_compps}). This  
seed-photon temperature, which is well below the energy range of the 
HXD, does not affect the spectral shape above 10 keV.
This model was found to provide an equally good fit 
(${\chi}^2/ \rm{d.o.f}=37/43$).  
The fit gave $\Gamma = 2.0 \pm 0.3$, which agrees with the 
power-law fit result, and the electron temperature was 
constrained only poorly as $kT_\mathrm{e} > 167$ keV, evidently 
because of the very flat shape of the hard tail.

\subsection{Analysis of the Broadband X-ray Spectrum}

We next analyzed the XIS and HXD spectra jointly. 
The cross-normalizations of PIN and GSO with 
respect to FI-XISs were fixed at 1.164\footnote{http://www.astro.isas.ac.jp/suzaku/doc/suzakumemo/suzakumemo-2008-06.pdf}, and that of 
BI-XIS was left free.
When using a convolution model,
we extended the energy band to 0.01--1000~keV.

\subsubsection{Simple Models}

The broad-band Suzaku spectra were fitted jointly with
a simple model consisting of 
a multi-color disk blackbody ({\tt diskbb}; \cite{Mitsuda1984}) plus 
a power-law model. The interstellar 
absorption was represented by {\tt tbabs} 
(\cite{Wilms2000}). 
However, when the photon index 
$\Gamma$ is varied, it became much higher ($\Gamma \sim 2.5$) than is indicated by 
the HXD data ($\Gamma = 2.0 \pm 0.4$, subsection~\ref{sec:fit_HXD}), 
and the fit failed with ${\chi}^2/ \rm{d.o.f}=843/472$. 
In particular, 
The HXD data deviated significantly from the model. 
This is because the XIS data have much better 
statistics than the HXD data, and the power-law component favored to fit the structure 
at the high energy end of the XIS spectrum (see below). 
We thus fixed $\Gamma$ at the value of 2.0 determined in 
subsection~\ref{sec:fit_HXD}.
This model was found to give a very poor fit, 
with ${\chi}^2/ \rm{d.o.f}=2682/468$. 
Figure~\ref{fig:figure_of_a_Suzaku_residual} 
(a) compares the data and the model in this unsuccessful fit, where
significant positive residuals remained around 5--10 keV.

This discrepancy could be caused by 
deformation of the disk emission spectrum due to  
relativistic effects, which are not considered 
in the {\tt diskbb} model. 
We assessed this possibility by replacing the 
{\tt diskbb} component to the {\tt kerrbb} model \citep{Li2005},  
which is a relativistic disk 
blackbody emission model for an accreting black 
hole with an arbitrary spin. Its input parameters 
are the spin parameter, the black hole mass, 
the inclination angle and distance of the system, 
the mass accretion rate, the color hardening factor, 
and another parameter $\eta$ which expresses 
the torque at the disk inner boundary. 
We fixed $\eta$ at 0 (corresponding to a 
torque-free inner edge), the distance 
at $8$ kpc, and the hardening factor at 1.7 (\cite{Shimura1995}), 
and let the other parameters vary freely.
However, as shown in figure~\ref{fig:figure_of_a_Suzaku_residual}(b),
the residual structure was not substantially mitigated from the 
{\tt diskbb+powerlaw} ($\Gamma =2.0$) fit, and the fit 
ended with ${\chi}^2/ \rm{d.o.f}=1619/472$.
Therefore, an additional spectral component is likely 
to be present at 5--10 keV.

\subsubsection{Single-Zone Comptonized Corona with Reflection}

An alternative explanation of the structure in 5--10 keV 
is a broad Fe-K$\alpha$ emission line produced 
when the Comptonized photons are reflected on the disk (e.g., \cite{Reis2010}). 
The emission line can be strongly broadened due to the relativistic 
effects, if the reflected photons come from a close 
vicinity of the black hole. To test if this is the case, we next added 
a relativistic reflection component 
to the {\tt diskbb} plus power-law model. 
Specifically, we adopted {\tt ireflect} and {\tt kdblur} as the reflection 
model and its relativistic smearing, respectively. 

The {\tt ireflect} model, which is a convolution version of the 
{\tt pexriv} model (\cite{Magdziarz1995}), calculates a reflection 
spectrum generated by ionized 
materials. Its free parameters are; the solid angle 
$\Omega$ of the reflector normalized by $2 \pi$, 
which parametrizes the reflection strength; 
the ionization parameter defined as $\xi = L_\mathrm{X}/(nR^{2})$, 
where $L_\mathrm{X}$, $n$, and $R$ are the ionizing luminosity 
in 0.005--20 keV, the density of the reflector, 
and the distance from the X-ray source; and the temperature 
of the reflector, $T_\mathrm{disk}$. 
The reflector was assumed to have  
solar abundances, and $T_\mathrm{disk}$ was fixed at 30000 K 
because it cannot be constrained by the Suzaku data. 
Because the {\tt ireflect} model does not account for 
any emission lines, we added a Gaussian component 
as the Fe-K$\alpha$ emission line. 
The normalization of Gaussian was linked with $\Omega /2 \pi$ of 
{\tt ireflect}, so that the line equivalent width with respect to 
the reflection continuum is always kept to be $\sim$1 keV, a typical 
value for reflection in black hole accretion disks (\cite{Matt1991}).
To reproduce the strong residual structure in 5--10 keV, 
we tested the case of $\Omega/2 \pi = 2$, in which the reflector 
covers the entire solid angle.
As the data favored high ionization as $\xi \sim 10^4$ erg cm s$^{-1}$, 
we fixed the line-center energy at 6.9 keV, 
to make it is consistent with the $\xi$ value.

The {\tt kdblur} model (\cite{Laor1991}) is a convolution model 
to smear, by relativistic effects, a spectrum generated in the 
inner part of accretion disks. It has 
four input parameters: the inner and outer radii of the 
accretion disk ($R_\mathrm{in}$ and $R_\mathrm{out}$, respectively) in units of 
$GM_\mathrm{BH}/c^2$, where $G$, $M_\mathrm{BH}$, and 
$c$ are the gravitational constant, the black hole mass, and the light speed, 
respectively; the inclination angle $i$; and the emissivity index $\beta$, which 
dictates that the line emissivity scales as $\propto R^{-\beta}$.
Together with $i = 60^\circ$ and $R_\mathrm{out} = 
400 R_\mathrm{g}$, we assumed $\beta = 3$, which is 
expected in a flat disk illuminated by a point source located 
in a finite distance above the black hole.
To investigate the case wherein the reflection spectrum 
is maximally smeared, we fixed 
$R_\mathrm{in} = 1.235$,
which implies an extreme Kerr black hole.

\begin{figure}[h]
\begin{center}
\FigureFile(76mm,76mm){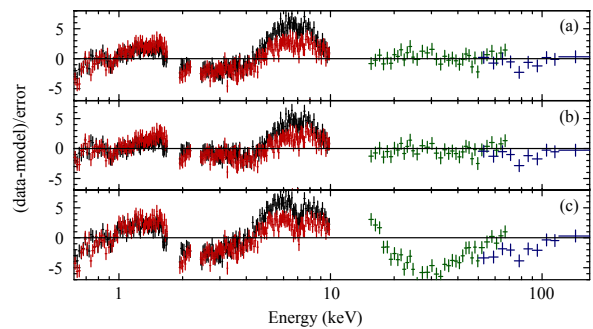}
\end{center}
\caption{(a--c) Ratios between the data and the individual models, 
using the same colors as in figure~ref{fig:fig3}: 
(a) {\tt tbabs*(diskbb+powerlaw)}, 
(b) {\tt tbabs*(kerrbb+powerlaw)}, and 
(c) {\tt tbabs*(diskbb+kdblur*(gauss+ireflect*powerlaw))}.
}
\label{fig:figure_of_a_Suzaku_residual}
\end{figure}

Combining all the components described above, 
we constracted a model expressed as 
{\tt tbabs*(diskbb+kdblur*(gauss+ireflect*powerlaw))},
and applied it to the data.  
However, as shown in 
figure~\ref{fig:figure_of_a_Suzaku_residual}(c) 
the model failed to explain
the observed structure in 5--10 keV, 
and the reflection continuum overestimated the 
X-ray flux around 20 keV. Thus, the obtained 
fit statistics, ${\chi}^2/\rm{d.o.f}=3988/472$, 
became much worse than those obtained with the 
simpler models tested above.

\subsubsection{Double Comptonization Model} \label{sec:double_compps}

Yet another possibility is that 
the observed structure is 
an additional Comptonization component, on top of 
the main power-law tail which dominates the hard X-ray 
flux above 10 keV.
Indeed, in some previous works of BHBs in the 
high/soft state, and the intermediate state with a 
strong disk blackbody component which is called the 
``soft intermediate state'', 
the X-ray spectra were successfully reproduced by 
two different Comptonization components
(e.g., \cite{Kubota2001}; \cite{Kolehmainen2011}; 
\cite{Nakahira2012}; \cite{Kawano2017}).
We adopted this ``double Comptonization'' model 
in an attempt to describe the residual structure in 5--10 keV, 
and considered two different geometries of 
the plasma responsible for this additional Comptonization: 
(1) it fully covers the standard disk  
(e.g., \cite{Nakahira2012}), and (2) it covers the disk only partially,  
and the remaining part of the disk is directly visible (e.g., \cite{Kawano2017}).

To test the case (1), we replaced, following \citet{Nakahira2012}, 
the direct MCD component in the MCD$+$power-law model to the 
Comptonization component {\tt compPS} \citep{Poutanen1996}. 
Using the numerical solutions of the radiative 
transfer equation, the {\tt compPS} model produces the Comptonized 
spectrum, in terms of the electron temperature, the optical depth $\tau$ of scattering, 
the energy distributions of incident photons
and electrons, and geometry of the electron cloud. 
We assumed the MCD emission as the 
seed photons, chose the ``slab'' geometry 
({\tt geom} = 1) for the electron cloud, 
and considered only thermal electrons ({\tt gmin} = 1). 
In this case, the normalization of {\tt compps} is 
defined as that of the seed spectrum (i.e., the 
disk blackbody spectrum), in the same way as {\tt diskbb}.  
This model can reproduce 
the reflection component, but we ignored it 
here by setting the solid angle of the 
reflector as $\Omega = 0$.

\begin{figure*}[h]
\begin{center}
\FigureFile(80mm,80mm){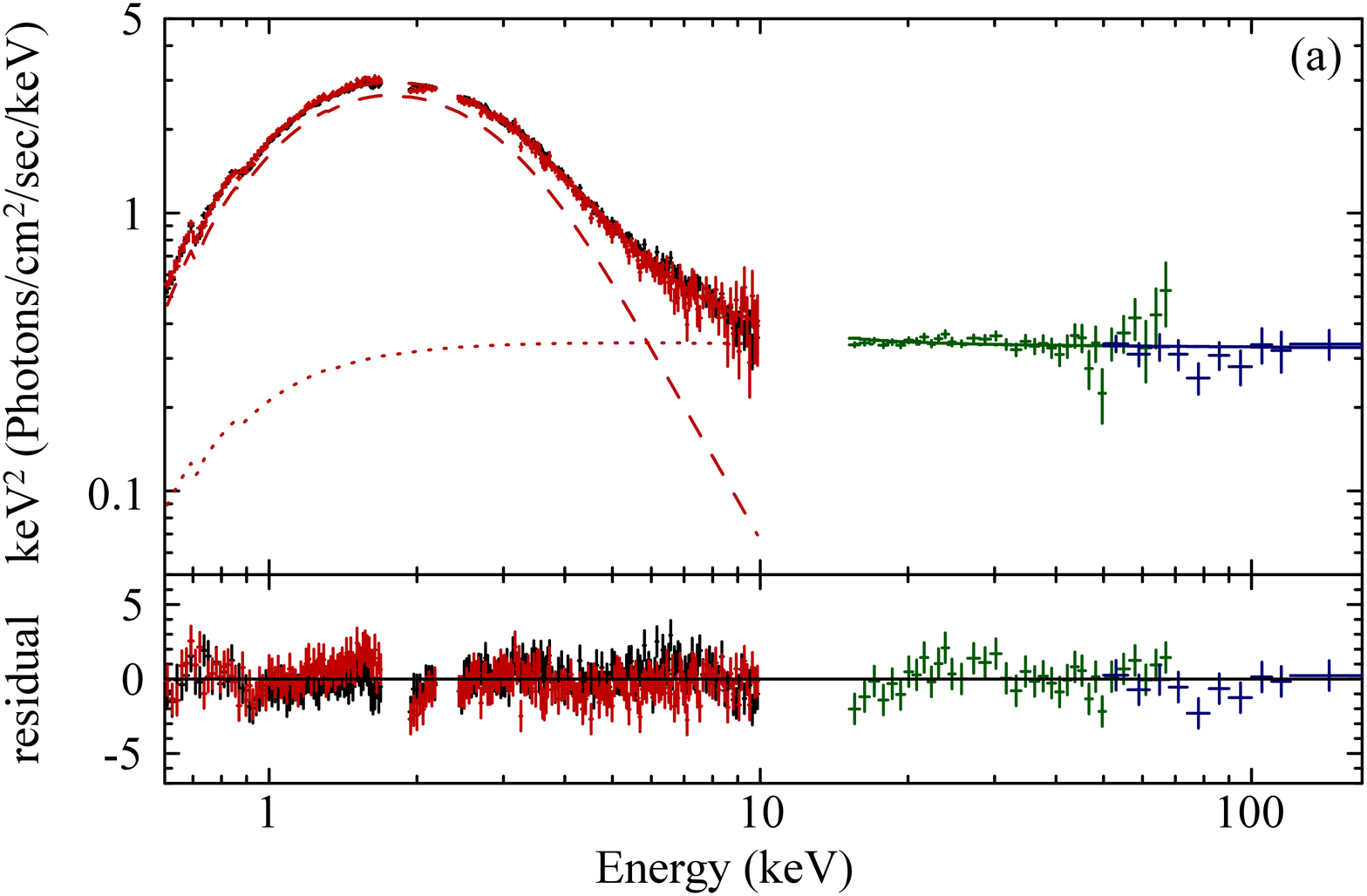}
\FigureFile(80mm,80mm){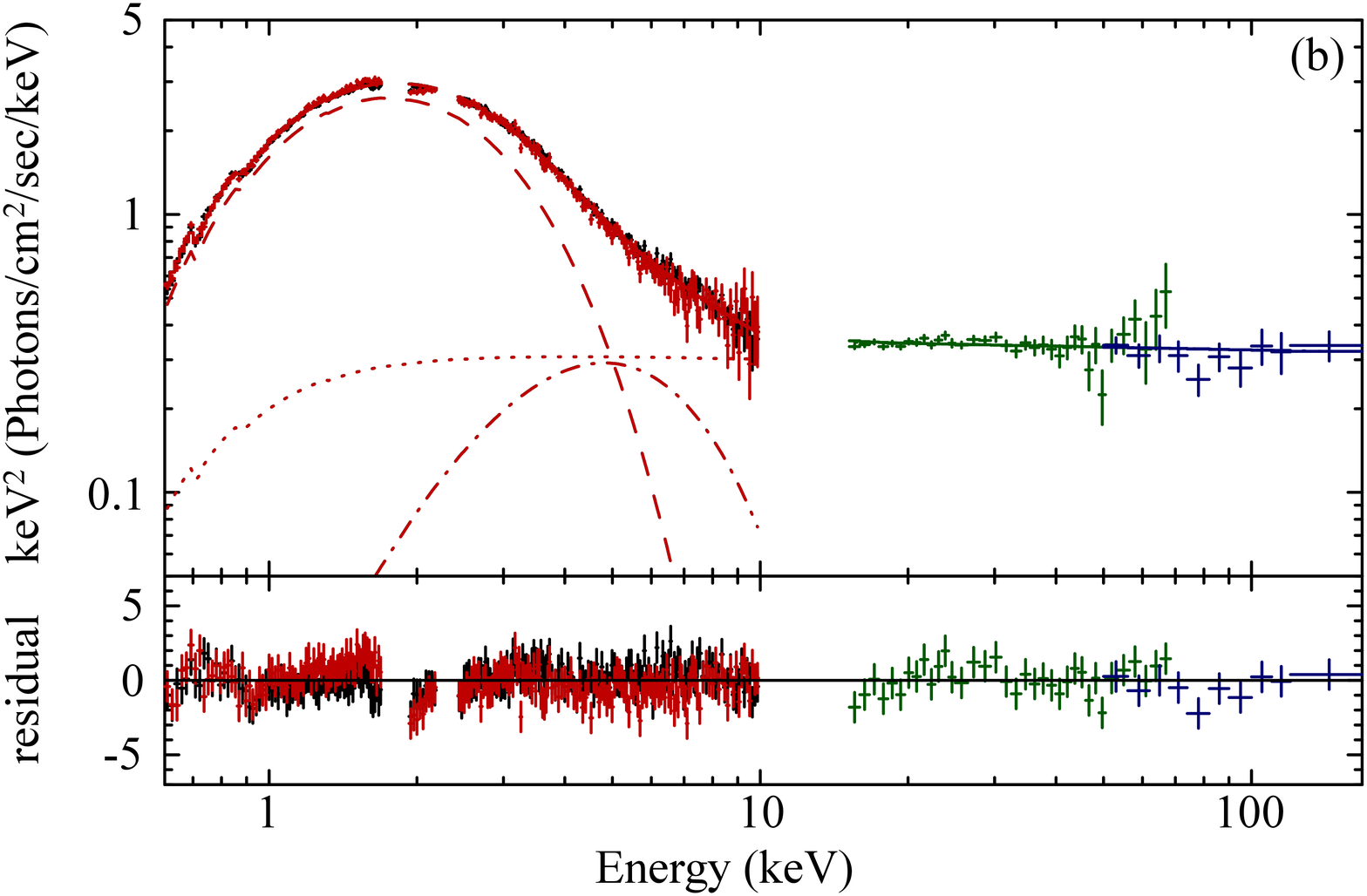}
\end{center}
\caption{Time-averaged Suzaku spectra of 
MAXI J1828$-$249, fitted with two successful double-Comptonization models, 
using the same colors as in figure~\ref{fig:fig3}.
(a) The fit with the {\tt tbabs*(compPS+powerlaw)} model, i.e., the 
case (1) modeling, where 
the dashed and dotted lines indicate the {\tt compPS} and {\tt powerlaw} 
components, respectively.  
(b) The fit with the {\tt tbabs*(diskbb+nthcomp+powerlaw)} model, i.e., the 
case (2) modeling, where
the dashed, dot-dashed, and dotted lines indicate 
the {\tt diskbb}, {\tt nthcomp}, and {\tt powerlaw} 
components, respectively. The lower panels show residuals between the data and the model.
}
\label{fig:figure_of_a_Suzaku_eeu_spectrum}
\end{figure*}

The {\tt compPS+powerlaw} model has successfully reproduced the spectrum, 
with $\chi^2$/d.o.f. $=$ 452/464 which is greatly improved over the 
MJD$+$power-law model. 
The best-fit parameters and the fit results 
are presented in figure~\ref{fig:figure_of_a_Suzaku_eeu_spectrum}{a} 
and table~\ref{tab:1}, respectively. 
Thus, the residual feature in 5--10 keV has disappeared. 
The {\tt compPS} parameters became $kT_\mathrm{e} = 14 \pm 7$ keV, which is 
relatively low, 
and $\tau = 1.0^{+1.1}_{-0.4}$.

Next, to test the case (2), 
we added an {\tt nthcomp} model 
to the MCD$+$power-law model, following \citet{Kawano2017}, 
and constructed a model as {\tt tbabs*(diskbb+powerlaw+nthcomp)}.
The reason for using {\tt nthcomp}, 
instead of {\tt compPS}, is that 
{\tt nthcomp} is suited to cases with $\tau \gtrsim 2$
and relatively low $kT_\mathrm{e}$ 
\citep{Zycki1999}, compared to {\tt compPS} which 
is appropriate for cases with $\tau \lesssim 1$ 
and high $kT_\mathrm{e}$.
The choice is justified later by the fit result. 
Again, the MCD spectrum was assumed as the seed 
spectrum for {\tt nthcomp}, and its innermost 
temperature was linked to that of the main MCD 
component. 
As shown in figure~~\ref{fig:figure_of_a_Suzaku_eeu_spectrum}(b) and 
table~\ref{tab:1}, this model has also given an acceptable fit, with 
${\chi}^2/ \rm{d.o.f}=440/463$, which is comparable to 
that of the {\tt compPS+powerlaw} fit.
To fit the structure at 5--10 keV, the {\tt nthcomp} model 
attained a low temperature as $kT_\mathrm{e} = 1.2^{+0.3}_{-0.05}$ keV, 
and a photon index of $\Gamma = 1.0^{+2.6}_{-1.0~\mathrm{pegged}}$.
The optical depth of the Comptonized cloud was estimated 
as $>3.5$, from $\Gamma$ and $kT_\mathrm{e}$ of the best-fit 
{\tt nthcomp} model, assuming a slab geometry, and using 
the equation in \citet{Zdziarski1996} and \citet{Hori2014}, as
\begin{equation}
\tau = \frac{1}{2} \left[ \sqrt{\frac{9}{4} + 
\frac{3}{\frac{kT_\mathrm{e}}{mc^2} 
\left(
\left( \Gamma +\frac{1}{2} \right)^2  
-\frac{9}{4} 
\right)}}  -\frac{3}{2} \right]. \label{eq:tau}
\end{equation} 
Therefore, the use of {\tt nthcomp} has been justified. This implies 
a much lower $kT_\mathrm{e}$ and a much higher $\tau$, compared with 
the case (1) solution. 

Based on the above two models, 
we also calculated the inner radius $R_\mathrm{in}$ of the 
standard disk, considering both the direct disk photons and the 
Compton-scattered photons. 
Assuming the conservation of the number of disk photons, 
there holds a relation as  
\begin{eqnarray}
\frac{{F_\mathrm{disk}}^\mathrm{p}}{2 \cos i} + {F_\mathrm{thc}}^\mathrm{p}  
= 0.0083 \left[ \frac{{R_\mathrm{in}}}{(D/ 10~\rm{kpc})^2} \right] 
\left( \frac{T_\mathrm{in}}{1~\rm{keV}} \right)^3 \nonumber \\~~ \rm{photons~s^{-1}~{cm}^{-2}},
\label{eq:eqEstimateOfRin}
\end{eqnarray}
where ${F_\mathrm{disk}}^\mathrm{p}$ and ${F_\mathrm{thc}}^\mathrm{p}$ 
represent the 0.01--100 keV photon flux in the observer's direction 
of 
the disk component and that of isotropic Comptonization components, 
respectively (\cite{Kubota2004}). 
This equation assumes that ${F_\mathrm{disk}}^\mathrm{p}$,
namely, the disk flux emitted towards the observer,
has an intrinsic emissivity which is proportional to $\cos i$,
and the denominator $2 \cos i$ is meant to correct
the observed ${F_\mathrm{disk}}^\mathrm{p}$ for this implicit effect.
As a result, the equation becomes rather inaccurate towards 
$i \rightarrow{} 90^\circ$. The factor of 2 is introduced simply
because $2 \cos i$ becomes unity when we take its spherical average.
The Comptonized photons are assumed to have rather weak anisotropy, if any.
Furthermore, this equation assumes that few photons are scattered
back to the disk, so that the corona-to-disk feedback effects are 
negligible. This condition is satisfied if the corona has a low optical 
depth. It is considered to apply to the present case, wherein 
$F_\mathrm{thc}^\mathrm{p}$ is dominated by the $\Gamma = 2$ power-law,
of which the source corona is likely to have a rather low optical 
depth from Equation~(\ref{eq:tau}), $\tau < 0.4$,
assuming a thermal Comptonization origin with $kT_\mathrm{e} > 169$ keV.

Our best-fit case (2) model gave $F_\mathrm{disk}^\mathrm{p} = 11.1$ 
photon s$^{-1}$ cm$^{-2}$ and ${F_\mathrm{thc}}^\mathrm{p} = 37.1$ 
photon s$^{-1}$ cm$^{-2}$,
which yield $R_\mathrm{in} = (1.2 \pm 0.2) 
\times 10^{2}~(\cos i/\cos 60^{\circ})^{-1/2}$ $(D/8~\mathrm{kpc})$ km.
By multiplying the correction factor 1.19 (e.g., \cite{Kubota1998}; 
\cite{Shidatsu2011}), which is based on the color hardening factor  
1.7 (\cite{Shimura1995}) and the stress-free inner boundary condition, 
the actual inner disk radius is estimated as 
$R_\mathrm{in}^* = (1.4 \pm 0.1) \times 10^{2}~(\cos i/\cos 60^{\circ})^{-1/2}$ $(D/8~\mathrm{kpc})$ km. 
In the case (1) model, we estimated 
$F_\mathrm{thc}^\mathrm{p}= 36.4$ photon s$^{-1}$ cm$^{^2}$ 
from the power-law component, and $F_\mathrm{disk}^\mathrm{p} = 11.6$
from the seed spectrum of the {\tt compPS} component, which was 
obtained by using {\tt diskbb} with its normalization 
set to the best-fit normalization of {\tt compPS}. 
These values gave, after multiplying the correction factor, 1.19, 
$R_\mathrm{in} = (1.5 \pm 0.1) \times 10^{2}~(\cos i/\cos 60^{\circ})^{-1/2}$ $(D/8~\mathrm{kpc})$ km, 
which is consistent with that obtained 
from the case (1) modeling. 
However, these values are still subject to 
various systematic uncertainties, because 
there can be other possible model variants besides the 
two cases considered here. 

\begin{table*}
\tbl{Best-fit parameters of the 
finally accepted two double-Comptonization models.}{%
\begin{tabular}{lccc}
\hline
Component&Parameter& Case (1) model\footnotemark[$*$] & Case (2) model\footnotemark[$\dag$]  \\
\hline
{\tt tbabs} & $N_\mathrm{H}(10^{22}~\rm{cm^{-2}})$ & $0.201 \pm 0.005$ & $0.194 \pm 0.004$\\
{\tt diskbb}\footnotemark[$\ddag$] & $kT_\mathrm{in}~(\rm{keV})$ & -  & $0.642_{-0.025}^{+0.006}$ \\
& norm & -  & $(2.6 \pm 0.1)\times 10^{3}$ \\
{\tt compPS}, {\tt nthcomp}\footnotemark[$\S$] & $kT_\mathrm{bb}~(\rm{keV})$ & $0.60_{-0.01}^{+0.02}$ & $=kT_\mathrm{in}$ \\
& $kT_\mathrm{e}~(\rm{keV})$ & $14 \pm 7$ & ${1.2}_{-0.1}^{+0.3}$ \\
& ${\Gamma}_\mathrm{th}$ & - & $1.0_{-1.0~\mathrm{pegged}}^{+2.6}$ \\
& $\tau$ & $1.0_{-0.4}^{+1.1}$ & $ >3.5$\footnotemark[$\parallel$] \\
& norm & $(3.4 \pm 0.3) \times 10^{3}$ & $1.4^{+17.8}_{-0.1} \times 10^{-2}$ \\ 
{\tt powerlaw} & ${\Gamma}$ & $2.01 \pm 0.04$ & $2.03 \pm 0.03$ \\ 
& norm & $0.35 \pm 0.05$ & $0.33_{-0.03}^{+0.04}$ \\
& ${\chi}^{2}/\rm{dof}$ & 451.69/464 & $440/463$ \\
\hline
\end{tabular}}\label{tab:1}
\begin{tabnote}
\footnotemark[$*$]{The {\tt tbabs*(compPS+powerlaw)} model, which assumes all 
disk photons are Comptonized into the {\tt compPS} component.}\\
\footnotemark[$\dag$]{The {\tt tbabs*(diskbb+nthcomp+powerlaw)} model, which assumes 
that only a fraction of disk photons are Comptonized, as represented by {\tt nthcomp}.}\\
\footnotemark[$\ddag$]{Not used in the case (1) model.}\\
\footnotemark[$\S$]{The seed spectrum was assumed as an MCD 
(i.e., {\tt inp\_type} was set to 1 for {\tt nthcomp} and 
the seed temperature to a negative value 
for {\tt compPS} in XSPEC). The seed temperature 
$kT_\mathrm{bb}$ was linked to $kT_\mathrm{in}$ 
of the {\tt diskbb} component in the case (2) model.} \\
\footnotemark[$\parallel$]{Calculated through Equation~(\ref{eq:tau}).} 
\end{tabnote}
\end{table*}

\subsection{Search for Ionized Absorption Lines}

The XIS spectra were searched for highly ionized 
absorption lines, originating e.g., from disk winds. 
To improve the photon statistics, we used  
the data extracted from the whole $108''$ circular region 
in the XIS image, without excluding the PSF 
core. However,  as shown in the inset of 
figure~\ref{fig:fig3}, no absorption lines were found 
around the energies of the Fe K lines. 
To determine the upper limits on the H-like 
and He-like Fe lines at 6.7 keV and 6.9 keV, 
respectively, which are usually the strongest lines in 
other BHBs with winds, we fitted the whole-region XIS 
spectrum again with the {\tt 
diskbb+nthcomp+powerlaw} model, 
plus two Gaussians representing the two lines. 
We fixed the line center energies of the Gaussians 
at 6.7 keV and 6.9 keV, and the line widths 
at 10 eV, 
which is much smaller than the energy resolution 
of the XIS. In addition to the two Gaussian normalizations 
which are assumed to be negative, 
we varied the normalizations of the continuum components  
and the photon index of {\tt powerlaw}, 
but fixed the other continuum parameters at the 
best-fit values obtained in section~\ref{sec:double_compps}.
From this analysis, the upper limits of 
the equivalent widths of H-like and He-like 
Fe lines were estimated to be $10$ eV 
and $9$ eV, respectively. These values 
are not expected to differ significantly, even if the lines are 
Doppler shifted as observed in some other BHBs (e.g., \cite{Miller2006}; \cite{Ueda2009}).

\section{Optical Observations and Results} \label{sec:OPTobs}

\subsection{Observation and Data Reduction}
As indicated in figure~\ref{fig:fig3}, optical photometric 
observations of MAXI J1828$-$249 
in the $R$ and $I$ bands were carried out on 2013 October 21 UT 9:25, 
simultaneously with the Suzaku observation.  
We used Hiroshima One-shot Wide field Polarimeter 
(HOWPol; \cite{Kawabata2008}), 
attached to the Kanata 1.5 m telescope at 
Higashi-Hiroshima Observatory, Hiroshima Astrophysical 
Science Center, Hiroshima University. 
On that night, 60 s $\times$ 9 $I$-band exposures were 
taken, followed by 60 s $\times$ 66 $R$-band exposures.

Using IRAF\footnote{http://iraf.noao.edu/}, we reduced the data 
in a standard way for optical imaging observations;  
bias subtraction and flat fielding were applied to the individual frames. 
The apparent magnitude of the target in each frame was estimated via 
PSF photometry, using 6 reference stars near the target in the field of view.
The magnitude error in each frame was calculated as standard deviation 
of the relative magnitudes derived using different reference 
stars. 
The apparent magnitudes in the $R$ and $I$ bands, averaged 
over the entire observation periods, were $\approx$17 mag and 
$\approx$16 mag, respectively,  
in the Vega unit.

We also used the optical and UV data (OBSID$=$00032997006) of 
the Swift UltraViolet and Optical Telescope (UVOT; \cite{Roming2005}) 
obtained on the same day, but 
$\sim$1 hour before the start of the Suzaku 
observation. Specifically, the UVOT observation was performed from 
2013 October 21 UT 04:25 to 04:42 through $UVW1$, $U$, $B$, $UVW2$, $V$, 
and $UVM2$ filters, with a net exposure of 85--342 s each. 
We defined the source area as a circle of a $5''$ radius, 
centered at the source position, 
and derived the source fluxes in the individual bands  
with a ftool {\tt uvot2pha}. The background levels were estimated 
from source-free regions. 

In addition to the above observations in the outburst, 
we searched archival data in quiescent phases for possible 
detections of the optical counterpart of MAXI J1828$-$249. 
The source was detected in the Pan-STARRS 
(Panoramic Survey Telescope and Rapid Response System) 
project in the $g$, $r$, $i$, $z$ and $y$ bands, 
on 2012 August 25. 
The $i$ and $r$-band apparent magnitudes were 18
and 19 (in the AB magnitude unit), respectively. 
When converted to the flux unit, these values imply 
that the object was about 4 times fainter 
than in our Kanata observations.

\subsection{Analysis of the Multi-Wavelength Spectral Energy Distribution} \label{sec:SEDfit}

\begin{figure}[h]
\begin{center}
\FigureFile(77mm, 77mm){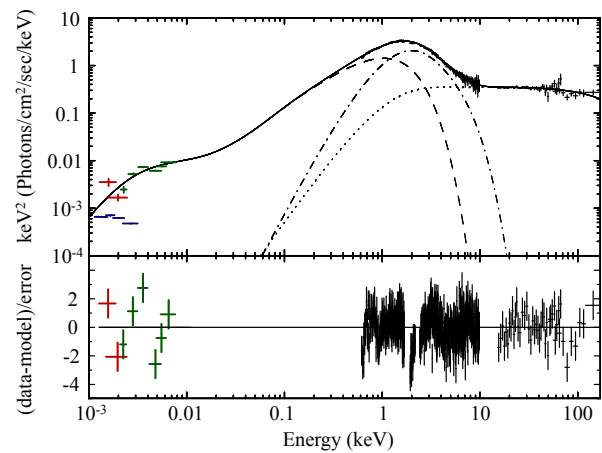}
\end{center}
\caption{Multi-wavelength SED of MAXI J1828$-$249, obtained with Suzaku (black), 
Kanata (red), and the Swift/UVOT (green), fitted with the {\tt optxrplir} model 
(black solid line). The Pan-STARRS data taken on 2012 August 25 
(before the outburst; not used in the SED fit}) are also plotted in blue.
The data and model are corrected for the interstellar absorption and extinction. 
The dot-dashed, dotted, and dashed lines indicate the contribution of 
the soft and hard Comptonization components, and the direct disk emission including 
the reprocessed component, respectively. The bottom panel shows the residuals 
between the data and the best-fit model.

\label{fig:fig8}
\end{figure}

Figure~\ref{fig:fig8} presents the quasi-simultaneous 
multi-wavelength spectral energy distribution (SED) 
of MAXI J1828$-$249, produced from the 
time-averaged data of Suzaku, Kanata, and Swift/UVOT. 
It is corrected 
for the X-ray absorption and the optical/UV extinction, 
using the model described below. The Pan-STARRS data are 
also shown in the same figure. Using these data, we 
investigated the origin of the optical/UV emission.

While the X-ray emission from BHBs is most likely generated in 
the inner region of the accretion flow, their optical photons 
are considered to originate potentially from multiple 
sources, including
synchrotron emission from jets, 
thermal emission from the outer disk region, and  
blackbody emission from the companion star.
In the present case, the jet emission is unlikely to 
contribute to the optical flux, because normally jets 
are not observed when the X-ray spectrum is dominated 
by the MCD emission 
(\cite{Fender2001}; \cite{Fender2001_2}; \cite{Fender2004}; 
\cite{Fender2004_2}). Indeed, the source was 
not detected in the radio observation performed 
3 days before our Suzaku observation 
\citep{Miller2013}, whereas detected after 
the source returned to a genuine low/hard state 
\citep{Corbel2014}, which took place 
$\sim$4 months later. 
The emission from the companion star is also 
unlikely to be a main contributor  
to the optical flux. In fact, the Pan-STARRS data taken 
before the outburst indicate that it contributes 
only up to $\sim 20$\% 
of the optical flux (figure~\ref{fig:fig8}). 
Hence, we infer that the optical 
emission of MAXI J1828$-$249 in the outburst was predominantly 
produced in the outer disk as thermal emission.

To confirm the above inference and investigate 
the outer disk conditions, we analyzed the quasi-simultaneous 
multi-wavelength SED 
obtained with Suzaku, Kanata, and Swift/UVOT, using a broad-band continuum 
SED model consisting of the disk emission and its 
Comptonization.
Normally, the outer disk emission of BHBs is significantly 
enhanced due to irradiation by the X-rays from the 
inner disk region (e.g., \cite{Gierliski2008}, \yearcite{Gierliski2009}),
and thus the optical and UV fluxes exceed the lower-frequency
extrapolation of the X-ray-determined MCD component. 
To take into account this effect, 
we adopted {\tt optxrplir} model (\cite{Shidatsu2016},
\cite{Kimura2018}) to fit the SED.

\begin{table}
\tbl{Best-fit parameters of the {\tt optxrplir} model.}{%
\begin{tabular}{lccc}
\hline
Component & Parameter & Value \\
\hline
{\tt redden} & $E(B-V)$\footnotemark[$*$] & $0.246 \pm 0.005$ \\
{\tt tbabs} & $N_\mathrm{H}~(10^{22}~\rm{cm^{-2}})$ & $0.142 \pm 0.003$ \\
{\tt optxrpilr} & $M_\mathrm{BH}~(\MO)$ & $4.0~(\rm{fixed})$ \\
& $D~(\rm{Mpc})$ & $0.008~(\rm{fixed})$ \\
& $\log(L/L_\mathrm{Edd})$ & $-0.840 \pm 0.002$ \\
& $a_\mathrm{star}$ & $0~(\rm{fixed})$ \\
& $R_\mathrm{cor}~(R_\mathrm{g})$ & $44 \pm 2$ \\
& $\log R_\mathrm{out}~(R_\mathrm{g})$ & $5.3 \pm 0.1$ \\
& $\tau$ & $6.0 \pm 0.6$ \\
& $R_\mathrm{pl}~(R_\mathrm{g})$ & $14.30_{-0.08}^{+0.09}$ \\
& ${\Gamma}_\mathrm{h}$ & $2.0~(\rm{fixed})$ \\
& $kT_\mathrm{eh}~(\rm{keV})$ & $1000~(\rm{fixed})$ \\
& $f_\mathrm{out}$ & $(3.2_{-0.8}^{+0.9}) \times 10^{-4}$ \\
& albedo & $0.3~(\rm{fixed})$ \\
& norm & $1.0~(\rm{fixed})$ \\
& ${\chi}^{2}/\rm{dof}$ & $539/472$ \\
\hline
\end{tabular}}\label{tab:2}
\begin{tabnote}
\footnotemark[$*$]{$E(B-V)$ is set to be $N_\mathrm{H}/(5.8 \times 10^{21}$ cm$^{-2}$).} 
\end{tabnote}
\end{table}

The {\tt optxrplir} model calculates a disk-emission spectrum 
and its Comptonization components, considering the irradiation  
of outer disk regions by the X-rays from the inner 
disk region. Like the predecessor SED models {\tt optxagnf} 
\citep{Done2012} and {\tt optxirr} \citep{Sutton2014},
it includes two different Comptonization components: a power-law tail 
computed based on the {\tt nthcomp} code (\cite{Zdziarski1996}; \cite{Zycki1999}) 
using the photon index $\Gamma_\mathrm{h}$ and 
an electron temperature $kT_\mathrm{eh}$ as free parameters, 
and a low-temperature thermal Comptonization component, which is based on the 
{\tt compTT} code \citep{Titarchuk1994} and is parametrized by 
the optical depth $\tau$ and another electron temperature $kT_\mathrm{es}$. 
The former is meant for the power-law tail generally seen 
in the high/soft state, and the latter for an additional 
lower-temperature Comptonization component, like the 
one which we found in the Suzaku spectrum. 
Here we set $kT_\mathrm{eh} =$ 1000 keV, because 
no significant exponential cutoff is present in the Suzaku 
energy band (figure~\ref{fig:figure_of_a_Suzaku_eeu_spectrum}b), 
and fixed $\Gamma_\mathrm{h} =$ 2.0, as 
obtained in section~\ref{sec:double_compps}. This model 
considers X-ray reprocessing effect, whose strength is 
determined by the height-to-radius ratio $f_\mathrm{out}$ 
of the outer disk, and 
the albedo $a_\mathrm{out}$. Because it is difficult to constrain 
the two parameters simultaneously, we fixed $a_\mathrm{out}$ at 0.3, 
following \citet{Shidatsu2016}, and allowed $f_\mathrm{out}$ to vary.
Further details of this model are given in Appendix 1.

The interstellar X-ray absorption and the 
optical/UV extinction were expressed by the {\tt tbabs} and 
{\tt redden} models, respectively. The {\tt redden} 
model calculates interstellar extinction with  
the reddening factor $E(B-V)$ as an input parameter, 
assuming the extinction factor $R_V = A_V/E(B-V) = 3.1$, 
where $A_V$ is the extinction measured in the $V$ band.
We varied $N_\mathrm{H}$ of {\tt tbabs} and linked it to $E(B-V)$ of {\tt redden}, 
so that they always satisfy the relation obtained by \citet{Bohlin1978}, 
$N_\mathrm{H}/E(B-V)=5.8 \times 10^{21}$ cm$^{-2}$ mag$^{-1}$.

This model was found to give an acceptable fit to the 
wide-band SED, with 
${\chi}^2/ \rm{d.o.f}=539/472$. The derived best-fit 
model is shown in figure~\ref{fig:fig8}, and its parameters 
are listed in table~\ref{tab:2}.
The irradiation strength, i.e., the fraction of 
X-rays reprocessed by the outer disk was obtained as 
$F_\mathrm{out} = 
(1-a_\mathrm{out}) f_\mathrm{out} = (2.3_{-0.6}^{+0.5}) \times 10^{-4}$. 
Because this is securely positive, the X-ray reprocessing effect is considered 
to have been significant in the outburst.
Indeed, as shown in figure~\ref{fig:fig8}, 
the optical and UV fluxes, corresponding to the outer 
disk emission, is strongly enhanced over the simple 
MCD extrapolations which would have a constant 
logarithmic slope toward lower energies. 

In figure~\ref{fig:fig8}, the MCD component and the 
soft Comptonization component 
add up to the soft X-ray spectrum in somewhat different ways 
than was obtained in subsubsection~\ref{sec:double_compps}.
This is mainly due to the difference 
of the modeling of the hard Comptonization 
component; 
the X-ray data analysis employed a power-law model 
which extends towards lower energies, whereas the 
SED analysis utilized {\tt nthcomp} which drops off 
below the seed photon peak.
However, the parameters related to the irradiation 
of the outer disk, $f_\mathrm{out}$ and $R_\mathrm{out}$ 
(see Appendix 1), which are the most important parameters in this SED analysis, 
are not significantly affected by 
this difference, because they are 
determined by the bolometric flux.

\section{Discussion} \label{sec:discussion}
\subsection{Summary of the Results}

Using Suzaku, we observed MAXI J1828$-$249 on 2013 October 21--22, 
and have successfully obtained broad-band X-ray data. 
It was $\sim$6 days after the MAXI discovery and the 
start of the spectral softening, and the source was in the 
brightest and softest 
condition ever achieved in the 2013 outburst.
Analyzing the Suzaku data, we have obtained the following results.
\begin{enumerate}
\item  
The rms variability of the 128-s binned light curve is
2--3\% in 0.6--5 keV (XIS), $\sim 5$\% in 5--10 keV 
(XIS) and 15--70 keV (HXD-PIN), and $<24$\% in 50--200 keV (HXD-GSO). 
\item  
No absorption lines are seen around the Fe-K energy region,
beyond an upper limit of $\sim$10 eV in equivalent width.
\item  
The 0.6--168 keV spectrum consists of a soft component,
and a clear hard-tail component
which extends at least to $\sim 170$ keV 
with $\Gamma \sim 2.0 \pm 0.4$ without bending.
The Comptonization modeling of the hard tail yields $kT_{\rm e}>167$ keV.
\item 
Although the continuum shape is typical of BHBs,
positive excess remains in 5--10 keV
when the spectrum is fitted with a simple MCD$+$power-law model.
The residuals cannot be explained either 
by relativistic broadening of the disk emission,
or by relativistic reflection which includes a broad Fe-K line.
\item 
The spectrum can be instead reproduced successfully
by adding a softer thermal Comptonization component,
assuming seed-photon supply from the disk.
The Comptonizing electron cloud covers the disk
either entirely with $kT_{\rm e} \sim 14$ keV and $\tau \sim 1$,
or partially with $kT_{\rm e} \sim 1.2$ keV and $\tau > 3.5$.
\item 
Regardless of the above ambiguity in the soft Compton cloud,
the innermost radius of the accretion disk is estimated as
$R_{\rm in} \sim 1.5 \times 10^2~(\cos i/\cos 60^{\circ})^{-1/2}$ $(D/8~\mathrm{kpc})$ km.
\end{enumerate}

Combining the present X-ray and optical data, 
with the two radio results (subsection 2.1), 
and the archival optical data in an X-ray quiescence, we have 
found that the optical emission of MAXI J1828$-$249 in our observations 
is likely to have been dominated by the irradiated outer disk emission, 
rather than the companion star or the jet emission. 
In fact, the multi-wavelength SED in the outburst was successfully described by 
a model consisting of the irradiated accretion disk emission and its 
Comptonization components.

\subsection{Identification of the Spectral State}

To identify the spectral state of MAXI J1828$-$249 in the 
Suzaku observation, we compare in table~\ref{tab:MAXIJ1828-HSS-LHS}
its characteristic parameters with those typically seen in the 
high/soft state and low/hard state. 
The overall shape of the broad-band spectrum, the inner disk 
temperature of $\sim$0.6 keV, and the power-law ($\Gamma \sim 2$) 
shaped hard tail without exponential cutoff, are all 
consistent with those of BHBs in the normal high/soft state. 
However, the intensity of the hard tail, relative to 
that of the MCD component, is somewhat large compared 
with what is usually seen in the high/soft state. In fact, 
as shown in figure~\ref{fig:figure_of_a_Suzaku_eeu_spectrum}a, the 
hard tail of MAXI J1828$-$249 is only $\sim$6 times fainter 
in the $E F_E$ form than the soft X-ray peak, 
whereas typical high/soft-state spectra have the hard tails 
which are $>$10 times fainter than their peaks (e.g., \cite{Done2007}).
Similarly strong hard tails are seen in other transient BHBs, 
such as GX 339$-$4 (e.g., \cite{Kolehmainen2011}; 
\cite{Plant2014}), Swift J1753.5$-$0127 (\cite{Chiang2010}) 
and MAXI J1910$-$057 (\cite{Nakahira2014}),
mainly during their state transitions. 
In reference to these knowledges accumulated so far, the present 
case of MAXI J1828$-$249, of which the power-law component 
contributes $\sim$40\% of the 2--15 keV flux, 
may be classified as soft intermediate state 
(\cite{Homan2005}; \cite{Plant2014}).

The soft Comptonization component, 
which is responsible for the 5--10 keV excess,
could be taken as another evidence for the intermediate state.
Actually, a similar soft Comptonization component is often seen
in Cygnus X-1, and it becomes enhanced in the intermediate state
(\cite{Yamada2013}; \cite{Kawano2017}). Similar examples include 
GX 339$-$4 \citep{Tamura2012} and XTE J1550$-$564 \citep{Gierlinski2003}.
However, this intermediate component is also observed in 
the high/soft state of some BHBs (\cite{Kolehmainen2011}; 
\cite{Shaw2016}). Therefore, it may be a preferred 
but not definitive feature of the intermediate state 
(see also subsection~\ref{sec:soft_compton}).

In addition to the spectral properties, the source 
has somewhat stronger variability, on time scales of 128 s to 
several tens ks, than transient BHBs in the high/soft state 
(table~\ref{tab:MAXIJ1828-HSS-LHS}).
Specifically, the estimated rms variability of 
MAXI J1828$-$249 in 5--10 keV, $\sim 5$\%, 
is larger than 
those in the typical high/soft state below $\sim$ 10 keV, 
$\sim 1$\% 
\citep{Heil2015}, but comparable to those of 
the intermediate state (e.g., \cite{Hori2014}). 
Further discussion continues in subsection~5.3.

Another interesting property of MAXI J1828$-$249 is 
its behavior on the hardness intensity 
diagram through the outburst, where no significant 
hysteresis is seen. This is reminiscent of that of 
MAXI J1836$-$194,
which exhibited an incomplete state transition and 
never reached the high/soft state.
Considering all the above results, and presuming that the 
hysteresis is observed only after the source completed 
its hard-to-soft state transition, we infer that 
MAXI J1828$-$249 stayed in the soft intermediate 
state and did not reach the genuine high/soft state 
in the Suzaku observation, as also suggested by
\citet{Filippova2014}. 
The observed properties of MAXI J1828$-$249, including the strong hard tail
and relatively enhanced variability above $\sim 5$ keV, are 
literally between those in the low/hard state and the high/soft state 
(table~\ref{tab:MAXIJ1828-HSS-LHS}). 
The soft intermediate state seems to be 
fairly stable at least in this object, 
as the hardness ratio stayed almost 
constant for $\sim$20 days around the Suzaku 
observation.

\begin{table*}
\tbl{Characteristic parameters of MAXI J1828$-$249 and of BHBs in the typical high/soft and low/hard states\footnotemark[$*$].}{%
\begin{tabular}{lccc}
\hline
Source & general & MAXI J1828$-$249 & general \\
State & high/soft state & soft intermediate & low/hard state \\
\hline
Disk component & & & \\
\hspace{0.5em} $F_\mathrm{disk}/F_\mathrm{total}$ (2--20 keV) & $> 0.75$ & 0.63 & $< 0.2$ \\
\hspace{0.5em} $kT_\mathrm{in}$ [keV] & 0.6-1.5 & 0.59-0.65 & $< 0.7$ \\
Hard tail & & & \\
\hspace{0.5em} $F_\mathrm{pl}/F_\mathrm{total}$ ($2$-$20~\rm{keV}$) & $<0.25$ & 0.37 & $> 0.8$ \\
\hspace{0.5em} $\Gamma$ & $> 2.1$ & 2.0-2.1 & 1.5-2.0 \\
Exponential cutoff & no & no & yes (at $\sim$100 keV) \\
Variability\footnotemark[$\dag$] & $\sim 0.01$\footnotemark[$\ddag$] & $\sim 0.05$\footnotemark[$\S$] & $\sim$0.1\footnotemark[$\ddag$] \\
Compact jets & no & no & yes  \\
\hline
\end{tabular}
}\label{tab:MAXIJ1828-HSS-LHS}
\begin{tabnote}
\footnotemark[$*$]{Mainly taken from \citet{McClintock2006} and this work, unless otherwise specified.}\\
\footnotemark[$\dag$]{Fractional rms variation in a frequency range of $5 \times 10^{-5}$--$8 \times 10^{-3}$ Hz}.\\
\footnotemark[$\ddag$]{In 2.5--13 keV, calculated by extrapolating the power spectra in 0.04--100 Hz 
given by \citet{Heil2015}.}\\
\footnotemark[$\S$]{In 5--10 keV.}
\end{tabnote}
\end{table*}

\subsection{Nature of the Additional Intermediate Component} \label{sec:soft_compton}

Here, we investigate the nature of the additional spectral structure found in 
5--10 keV. This is considered to be a relatively stable feature of 
MAXI J1828$-$249, as it is also present in the simultaneous
Swift and INTEGRAL spectra \citep{Grebenev2016}, taken on 
October 15--18 which are several days before our Suzaku observation.
This component is likely to be mildly variable, 
because the rms variability in 5--10 keV ($\sim 5$\%) 
is larger than that in 0.6--5 keV where the raw disk 
emission is seen (figure 6), and is comparable to that 
in 15--70 keV where the hard tail dominates.
Although the varying hard tail would add to the 5--10 keV 
variability, this can explain the observed 5--10 keV variability 
at most only partially, because the hard-tail photons 
contribute only $\sim 50$\% 
in this energy band.

The 5--10 keV structure in the Suzaku data has been reproduced
equally well by the two variants of Comptonization.
One (case 1) assumes that the entire disk region emitting X-rays
is covered by a relatively thin ($\tau \sim 1$) Comptonizing
electron cloud with $kT_{\rm e} = 14 \pm 7$ keV,
so that the original MCD emission as a whole is modified
into a Comptonized spectrum with a higher color temperature.
The other (case 2) is a condition wherein a thick ($\tau > 3.5$)
and much cooler ( $kT_{\rm e} \sim 1.2$ keV) Compton cloud 
covers a part of the disk, and the produced Wien hump 
peaked at $4 kT_{\rm e} \sim 5$ keV 
is added to the direct MCD emission.

The above two alternative solutions can be
distinguished by a parameter  $Q \equiv T_{\rm e}/T_{\rm in}$.
First introduced by \citet{Makishima2014} 
and applied to various classes of accreting X-ray sources
(\cite{Zhang2016}; \cite{Kobayashi2017}), 
it serves as a good indicator of the balance between
Compton cooling and ionic heating of the electrons.
The case 2 modeling yields $Q = 2$,
which means that the electrons are strongly cooled to
form a ``cool and thick'' Compton cloud.
Such low values as $Q < 7$ are often found also with the
Comptonized blackbody radiation of neutron-star low-mass 
X-ray binaries in their high/soft state 
(\cite{Zhang2016}; \cite{Makishima2014}).
If adopting the case 1 modeling instead, we obtain $Q=30$,
which is comparable to the values of $Q \sim 50$
found in so-called very high state and hard 
intermediate state of BHBs (\cite{Tamura2012}; \cite{Hori2014}).
Even higher values as $Q = 10^{2}-10^{3}$, to be called ``hot and thin'' 
Comptonization, are obtained from
the hard X-ray continuum of BHBs in the low/hard state
\citep{Makishima2008}.
Thus, the soft Compton component detected from MAXI~J1828$-$249
is more strongly cooled than the hot inner accretion flows 
in the low/hard state where the Comptonized hard continuum is produced,
and becomes comparable to those in the very high state (case 1 modeling),
or even further to reach a ``cool and thick'' condition (case 2 modeling).

So far, cool Comptonization components, similar to the present one,
have also been observed from some other BHBs, such as XTE J1752$-$223 
\citep{Nakahira2012}, Cygnus X-1 \citep{Kawano2017}, 
and GX 339$-$4 \citep{Kolehmainen2011}, 
when they showed disk dominant spectra. 
Therefore, this is neither specific to MAXI J1828$-$249,
nor a rare phenomenon. Nevertheless, the origin of these 
soft Compton X-rays is still unclear, and leaves us 
with at least two alternatives corresponding to the 
case (1)/(2) ambiguity. If adopting the case (2) standpoint, 
a likely scenario is that the standard disk is still 
developing toward the ISCO,
and the thick and cool Compton cloud is formed in a region
where the disk inner edge intrudes into the hot inner flow
(e.g., \cite{Kawano2017}).
Alternatively, we had better employ the case (1) scenario,
when the low-temperature Compton component co-exists 
with a standard disk that is likely to extend down to 
the ISCO \citep{Kolehmainen2011}. 
Then, the color hardening effect may be produced by 
a Comptonizing layer which somehow develops on the disk 
surface (\cite{Kolehmainen2011}; \cite{Nakahira2012}).

The energy dependent variability in figure~\ref{fig:rms}
may help us to distinguish the case (1)/case (2) degeneracy. 
In the case (2) modeling  (figure~\ref{fig:figure_of_a_Suzaku_eeu_spectrum}b), the emission 
below $\sim 5$ keV would be dominated by the disk photons, which 
are expected to vary only $\sim$ 1\% (table~\ref{tab:MAXIJ1828-HSS-LHS}). 
In contrast, the case (1) fit  (figure~\ref{fig:figure_of_a_Suzaku_eeu_spectrum}a) implies that 
the entire disk spectrum is Comptonized, so that the signals below 
$\sim$ 5 keV should be still variable as the corona varies 
in its optical depth and/or the electron temperature. Thus, the 
case (1) interpretation would be somewhat favored in MAXI J1828$-$249.

Whatever the origin of the low-temperature Comptonization
component in MAXI J1828$-$249 is, our result suggests that
the multi-zone Comptonization is ubiquitous among various 
states in BHBs, supposing that the hard tail in the disk 
dominated states is also produced by Comptonization of 
the disk photons. In fact, Suzaku observations have shown 
that a double Comptonization modeling is often required 
to reproduce BHB spectra in the low/hard state
(\cite{Makishima2008}; \cite{Takahashi2008}; 
\cite{Shidatsu2011}; \cite{Yamada2013}),
as well as in the very high state (\cite{Tamura2012}; 
\cite{Hori2014}).
Recent studies of short-term spectral variations
above $\sim 0.1$ Hz in the low/hard state
also support the multi-zone Comptonization picture
(\cite{Axelsson2018}; \cite{Mahmoud2018}).
Such a view is likely to apply also to Syefert galaxies, of 
which so-called soft excess component seen in $<2$ keV, 
is considered to arise as a soft Comptonization component 
(e.g., \cite{Noda2013}). 
In contrast, a single-zone Comptonization with $Q>7$ and $Q<7$
are usually sufficient to describe the low/hard state and  
high/soft state spectra, respectively, of low-mass 
X-ray binaries involving 
weakly-magnetized neutron stars (\cite{Sakurai2014}). 
Therefore, a clue to the problem may be obtained through
unified studies of these phenomena in various classes of 
accreting objects.

\subsection{Properties of the Hard Tail}

Another important result from the present Suzaku observation
is the clear detection of the hard tail, which extends 
from $\sim 15$ keV to $\sim 170$ keV with a very constant 
slope of $\Gamma = 2.0$ (figure 6). As summarized in table 3,
this flatness without high-energy bending makes it more similar 
to the hard-tail component seen in the high/soft state,
than to the low/hard state continuum.
Furthermore, the fractional rms variability of the hard tail,
$\sim 5$\% which we measured in 15--70 keV,
is similar to those of the hard tail in the high/soft state;
in fact, the 10--20 keV signals of Cygnus X-1 in the high/soft 
state are estimated to vary by $\sim 6$\%, as derived by 
integrating its long-term 10--20 keV power spectrum, 
obtained with the MAXI/GSC (figure 3 right of \cite{Sugimoto2016}), 
over $10^{-2}$ Hz to $10^{-5}$ Hz
which matches with figure 3.

In contrast to the above similarities, some dissimilarities are also 
noticed when the hard tail of MAXI J1828$-$249 
is examined from several other aspects (see also table 3).
These viewpoints include the rather high hard-tail luminosity relative to 
the disk luminosity, which we used as the primary feature in our state 
identification in subsection 5.2, and the value of $\Gamma = 2.0$
which indeed falls on the flattest end and the steepest end
of the $\Gamma$ distributions in the high/soft and low/hard states, 
respectively.
These two properties make the present hard tail literally intermediate 
between those seen in the two major states. 
We hence speculate that 
the hard X-ray continuum in the low/hard state
and the hard tail component in the high/soft state 
are related to each other in one way or another. The hard tail 
observed from MAXI J1828$-$249 is possibly a transition 
state, through which the former changes into the latter.

As described in Section~\ref{sec:soft_compton}, 
we consider Comptonization of disk photons in an 
electron cloud inside or above the disk, as a plausible 
origin of the hard tail of MAXI J1828$-$249. The electrons 
may have a non-thermal distribution or a thermal 
distribution with a temperature of $> 167$ keV. 
Some previous works proposed Comptonization in jets 
especially in the low/hard state and 
the hard intermediate state (e.g., \cite{Reig2003}; 
\cite{Markoff2005}, \cite{Reig2015}). 
Although this picture could explain the X-ray emission 
at the very onset of the outburst before the spectral 
softening, it is unlikely to give a satisfactory 
explanation of the hard tail seen in the Suzaku 
observation. In fact, the radio non-detection 
\citep{Miller2013} suggests that the jets already 
ceased a few days before 
that epoch. Multi-wavelength studies in various 
states would be needed to 
understand how much jets can contribute to the 
Comptonization components and how its contribution 
changes during the state transitions.

\subsection{Implication from the Multi-wavelength SED}

The optical and X-ray SED in the outburst was successfully 
reproduced with the {\tt optxrplir} model, one of the latest spectral 
models of an irradiated accretion disk. 
The outer disk radius was obtained as $R_\mathrm{out} = 10^{5.3} R_\mathrm{g}$, 
which is comparable with those in other transient BHBs with relatively 
short orbital periods ($\lesssim$ a few days), such as 
MAXI J1305$-$704 (\cite{Shidatsu2013}), MAXI J1910$-$057 (\cite{Nakahira2014}; \cite{Degenaar2014}), 
and GRO J1655$-$40 (e.g., \cite{Shidatsu2016}), but smaller 
than those with longer orbital periods, such as GRS 1915$+$105 
(e.g., \cite{Frank1992}; \cite{Done2004}) and V404 Cygni \citep{Kimura2016}.
The estimated value, $R_\mathrm{out} \sim 5$ $(M_\mathrm{BH}/5 \MO)$ 
lt-sec is comparable to the Roche-lobe radius of $\sim 2$ lt-sec, 
expected when a black hole with $5 \MO$ forms a binary with a 
main-sequence low-mass star of $\sim 0.5 \MO$ 
which marginally fills its Roche lobe. 

The strength of reprocessing, $F_\mathrm{out} = f_\mathrm{out} * (1-a_\mathrm{out})  
\sim 2.3 \times 10^{-4}$, is $\gtrsim$1 order of magnitude 
lower than the typical values in the high/soft and low/hard states 
(e.g.,~\cite{Gierliski2009}; \cite{Chiang2010};
\cite{Rahoui2012}; \cite{Shidatsu2013}; \cite{Nakahira2014}; \cite{Shidatsu2016}; 
\cite{Kimura2018}). 
This value could be even lower when the contribution 
of the companion star to the optical and UV fluxes is subtracted.
What made the small reprocessed fraction is still unclear, 
but it could be ascribed to a difference in the geometry of 
the outer disk, and/or the absorption efficiency on the disk 
surface. Yet another possibility is that the inner disk is fully 
covered by the cool Compton cloud (case 1 in subsection 3.3),
which is relatively optically thin ($\tau \sim 1$) as seen from above
but becomes optically thicker towards directions 
which are grazing to the disk plane. As a result, the emerging 
soft X-rays (with the Comptonized color) are mildly collimated 
in the directions perpendicular to the disk,
just like the limb darkening effect seen in the Solar photosphere.
If this interpretation is correct, the case (1) double Comptonization 
would be more favored, in agreement with the inference from the variability (subsection 5.3).

\begin{figure*}[h]
\begin{center}
\FigureFile(140mm, 140mm){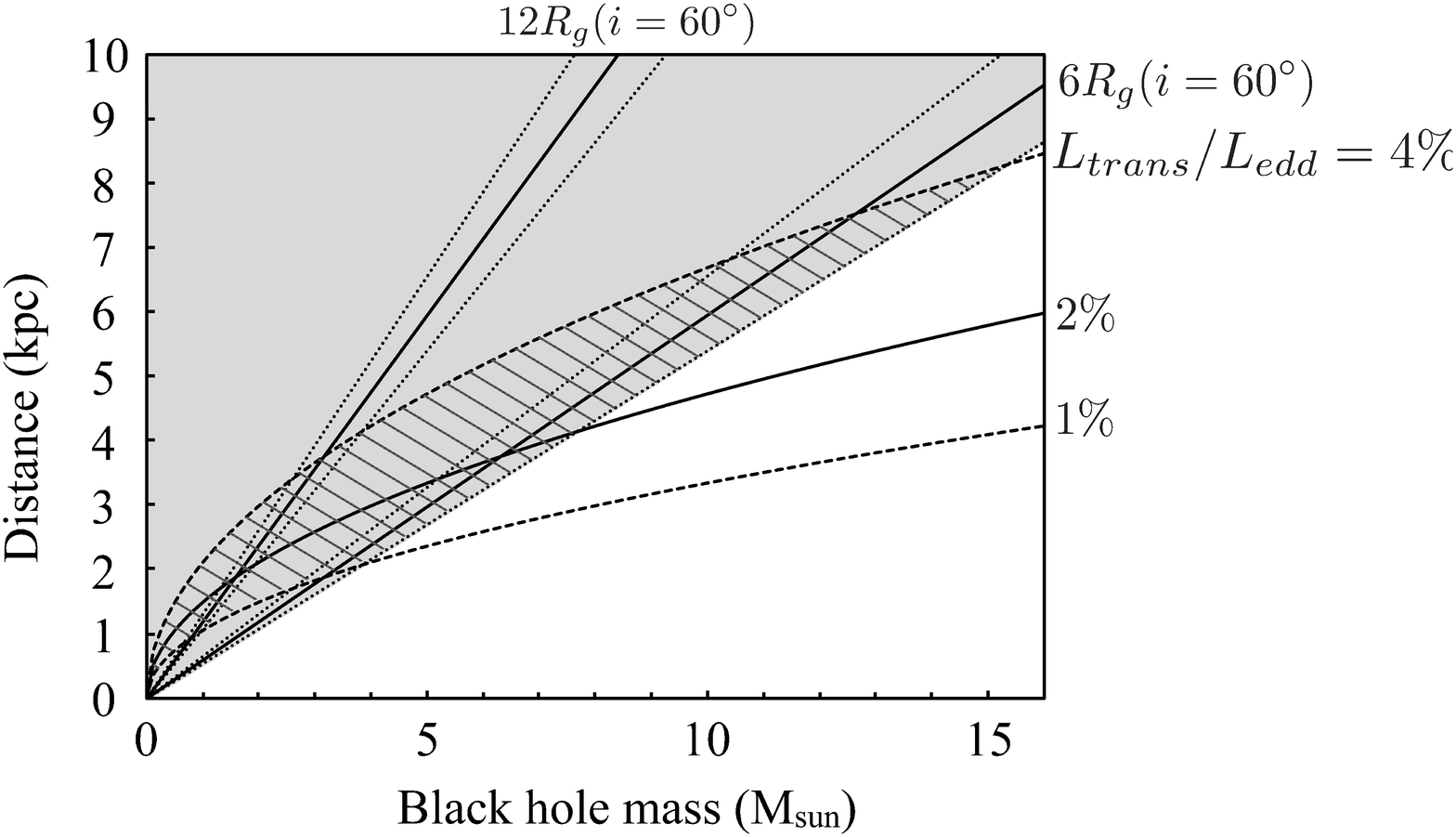}
\end{center}
\caption{Constraints on the black hole mass and distance of 
MAXI J1828$-$249. Solid lines indicates $R_\mathrm{in} = 6 R_\mathrm{g}$ 
and $12 R_\mathrm{g}$, which correspond to $R_\mathrm{ISCO}$ and $2 R_\mathrm{ISCO}$, 
respectively, assuming $i=60^\circ$ and a non-rotating black hole.
Dashed lines show their 90\% confidence limits.
The shadowed region satisfies $R_\mathrm{in} \geq R_\mathrm{ISCO}$.
Solid curves are obtained from the condition that the %soft-to-hard 
transition back to the low/hard state took place at 
1\%, 2\%, and 4\% of the Eddington luminosity, $L_\mathrm{Edd}$.
The hatched region satisfies $R_\mathrm{in} \geq R_\mathrm{ISCO}$ and 
the soft-to-hard transition luminosity $L_\mathrm{trans} =$
1\%--4\% $L_\mathrm{Edd}$.}
\label{fig:fig10}
\end{figure*}

\subsection{System Parameters of MAXI J1828$-$249}
Considering our X-ray, optical, 
and UV results, let us estimate basic parameters 
of this binary system, including the distance $D$ and the black hole mass 
$M_\mathrm{BH}$. The shadowed region in figure~\ref{fig:fig10}
indicates the allowed ranges on the $D$-$M_\mathrm{BH}$ 
space, obtained from the condition that $R_\mathrm{in}$ 
estimated from the best-fit double 
Comptonization model (see subsubsection~\ref{sec:double_compps}) 
is not smaller than the ISCO radius, $R_\mathrm{ISCO}$.
The solid lines, derived assuming a non-spinning black hole
(i.e., $R_\mathrm{ISCO} = 6 R_\mathrm{g}$) 
with $R_\mathrm{in} = R_\mathrm{ISCO}$ and 
an inclination angle of $60^\circ$, 
move downward, if a lower inclination 
is assumed. 
If the black hole is spinning rapidly, both the 
intensity and shape of the spectrum are affected 
in complex ways by various relativistic effects, 
including the decrease in $R_\mathrm{ISCO}$, light bending, 
gravitational redshifts, and Doppler beaming (see e.g., \cite{Wang2018}).
Further investigation using relativistic disk emission 
models, however, is beyond the scope of 
this work and we left as a future work. 

Figure~\ref{fig:fig10} also plots the $D$-$M_\mathrm{BH}$ 
relations, derived by estimating the Eddington luminosity 
of the objects from the actually observed luminosities. 
Even though MAXI J1828$-$249 may have stayed in 
the soft intermediate state without reaching the genuine 
high/soft state, the MAXI/GSC hardness ratio in 
figure~\ref{fig:fig1}c increased significantly over 
MJD 56619--56640, while the 15--50 keV Swift/BAT 
intensity remaining rather constant.
We therefore accumulated the MAXI/GSC data taken over 
this period into a single spectrum, and fitted it successfully 
with a power-law model with a photon index of $\sim 2.4$. The 
obtained 0.1--100 keV flux, $F = 
9.3 \times 10^{-9}$ erg s$^{-1}$ cm$^{-2}$, may be 
taken as the value when the system fully returned to the low/hard state. 
The three curves in figure~\ref{fig:fig10} were derived by equating 
this $F$ with 1\%--4\% of the Eddington luminosity, 
as normally observed in BHBs \citep{Maccarone2003}.
The hatched region indicates where the two 
different $D$-$M_\mathrm{BH}$ relations are satisfied, 
assuming $i=60^\circ$.

The exact location of the inner edge of the standard 
disk is unclear if the source is in the soft 
intermediate state. Although $R_\mathrm{in} \sim 
R_\mathrm{ISCO}$ is a likely condition, 
previous studies suggest that the standard disk is 
slightly truncated at less than a few times the 
ISCO in the very high state
(\cite{Tamura2012}; \cite{Hori2014}).
A consistency check of figure~\ref{fig:fig10} may be 
provided by the $N_\mathrm{H}$ value of MAXI J1828$-$249, 
(1.5--2.0) $\times 10^{21}$ cm$^{-2}$, which was obtained by 
analyzing the X-ray spectrum and the multi-wavelength SED. 
This is comparable to the total Galactic column, 
$N_\mathrm{H} = 1.6 \times 10^{21}$ cm$^{-2}$, and thus 
the source would be located near the Galactic center or 
farther. All the available information is consistent 
with an interpretation that MAXI J1828$-$249 is a binary 
consisting of a stellar mass black hole and a low-mass companion, 
and is located as a distance of several to $\sim$7 kpc.

The nature of the companion star is still unknown. 
Assuming a distance of 8 kpc, 
the Pan-STARRS data obtained before the outburst yields 
an absolute $V$ magnitude of $M_V \sim 3$. 
If the companion star is a main sequence star, 
it should be an $A \sim F$-type
or later type star (\cite{Turon1992}; \cite{Turon1998}).

\section{Summary and Conclusion} \label{sec:summary}

We observed the BHB candidate MAXI J1828$-$249 in X-ray, UV, and optical 
bands with Suzaku, Swift/UVOT and Kanata, respectively, when it reached 
its peak flux in the 2013 outburst. The main conclusions are summarized as follows.

\begin{enumerate}
\item The X-ray properties observed with Suzaku, including 
the somewhat stronger variability 
and the brighter hard tail than those of typical 
transient BHBs in the high/soft state, and the non-hysteretic 
behavior on the hardness intensity diagram, suggest that the source 
was in the soft intermediate state, rather than the high/soft state.
\item In the X-ray spectrum, a low-temperature Comptonization 
component is required in addition to the MCD component and 
the hard tail dominating above $\sim$ 10 keV. 
This spectral component can be equally well reproduced by 
assuming that this additional Comptonizing cloud 
covers the disk either partially or fully. 
The latter may be favored, considering the energy dependence 
of the rms variability.
\item The hard tail has transitional properties between 
those in the typical high/soft state and the 
continuum in the typical low/hard state. This 
suggest that they are somehow connected. 
\item A dominant fraction of the optical and UV emission 
was likely to be produced by the outer disk, which is relatively 
weakly irradiated by X-rays.
\item Constraints on the basic parameters of the binary system, 
including the black hole mass, distance, and the type of the 
companion star, were acquired. MAXI J1828$-$249 is considered as 
a binary containing a stellar-mass black hole.

\end{enumerate}

\begin{ack}
We are grateful to the Suzaku operation team for carrying out the ToO 
observation. 
This research has made use of MAXI data provided by RIKEN, JAXA and
the MAXI team. Part of this work was financially supported by 
Grants-in-Aid for Scientific Research 16K17672 (MS) and 16K05301 (HN) 
from the Ministry of Education, Culture, Sports, Science and Technology 
(MEXT) of Japan. MS acknowledges support by the Special Postdoctoral 
Researchers Program at RIKEN.
\end{ack}

\appendix 
\section{The optxrplir model}
Here, we describe the details of the SED model {\tt oprxrplir},  
utilized in section~\ref{sec:SEDfit}. 
It assumes the Novikov-Thorne emissivity profile 
\citep{Novikov1973}, over the entire region of the 
accretion disk from the ISCO to the outer disk edge. 
The ISCO radius is determined by the two 
input parameters: the black hole mass $M_\mathrm{BH}$ 
and the spin parameter $a_*$, and the outer disk 
radius is by $R_\mathrm{out}$, which specifies the 
outer disk radius in units of $R_\mathrm{g}$. 
In this model, the direct disk component and the hard and soft 
Comptonization components 
which are calculated on the basis of {\tt nthcomp} and {\tt compTT}, 
respectively, are separated radially; their luminosities are 
defined as the luminosities of the relativistic standard disk 
from $R_\mathrm{ISCO}$ to $R_\mathrm{pl}$, 
$R_\mathrm{pl}$ to $R_\mathrm{cor}$, and 
$R_\mathrm{cor}$ to $R_\mathrm{out}$, respectively. 
Thus, the input parameter $R_\mathrm{pl}$ determines 
the normalization of the hard Comptonization component, 
and $R_\mathrm{cor}$ determines that of the soft 
Comptonization components. 
The color-temperature correction is considered for the direct 
disk spectrum produced in the region above $R_\mathrm{cor}$, 
in the same manner as {\tt optxagnf} \citep{Done2012}.
The luminosity of each component is converted to the flux 
at the observer's position by using the distance $D$.

To account for the X-ray reprocessing component, {\tt optxrplir} 
assumes that the disk height $H(r)$ at a radius $r$ is proportional to $r^{9/7}$
and that the illuminating X-ray flux has a radial dependence 
of $r^{-12/7}$, unlike previous models such as {\tt diskir} 
(\cite{Gierliski2008}, \yearcite{Gierliski2009}) 
and {\tt optxirr}, which assume an $r^{-2}$ dependence 
for the illuminating flux. 
The radial dependence adopted in {\tt optxrplir} is 
based on a theoretical calculation of the irradiated 
geometrically-thin accretion disk (\cite{Cunningham1976}). 
To determine the strength of the reprocessed component, 
the {\tt optxrplir} model uses the albedo ($a_\mathrm{out}$) 
and the geometry-dependent factor ($f_\mathrm{out}$), 
as input parameters. The albedo is assumed to be constant radially, 
and therefore a fraction $(1-a_\mathrm{out})$ of the total incident X-ray 
flux are reprocessed at each radius. The parameter $f_\mathrm{out}$ 
is defined as the ratio of the disk height at the outer edge 
over the outer radius, and determines the fraction of 
the X-rays intercepted by the outer disk. 
The strength of the reprocessed component is thus 
scaled with $F_\mathrm{out} = (1-a_\mathrm{out}) f_\mathrm{out}$. 

%\begin{thebibliography}{}
\bibliographystyle{apj}
\bibliography{MAXI_J1828-249}
%\end{thebibliography}

\end{document}